\documentclass[12pt]{article}
\usepackage[utf8]{inputenc}
\usepackage[left=2.5cm, right=2.5cm]{geometry}
\usepackage[english]{babel}
\usepackage{pdfpages}
\usepackage{natbib}
\usepackage[backref=page]{hyperref}
\usepackage[titletoc,title]{appendix}
\usepackage{float}
\usepackage[short]{optidef}
\usepackage{setspace}
\usepackage{amsmath}
\usepackage{amssymb}
\usepackage{booktabs}
\usepackage{tabularx}
\usepackage{adjustbox}
\usepackage[calc]{datetime2}
\usepackage{graphicx}
\usepackage{listings}
\usepackage{soul}
\usepackage{authblk}
\usepackage{subcaption}
\usepackage{tcolorbox}

\usepackage{multirow}
\usepackage{fancyhdr}
\graphicspath{ {./pics/} }

\title{\color{black}{Nowcasting the euro area with social media data}}
\author[1]{Konstantin Boss}
\author[1]{Luigi Longo}
\author[1]{Luca Onorante\thanks{We thank our consultants, Florian Huber and Michael Pfarrhofer, and participants of the Conference ECONDAT 2025 spring meeting: ``Economics with nontraditional data and analytical tools'', for useful input. The views expressed herein are those of the authors and do not necessarily represent the views of the European Commission. All remaining errors are our own. Corresponding author: Luigi Longo, Via E. Fermi, 2749, 21027 Ispra (VA). Email: luigi.longo@ec.europa.eu.}}
\affil[1]{JRC - European Commission}

\date{\today}

\begin{document}
\doublespacing
\renewcommand\thetable{\arabic{section}.\arabic{table}}
\renewcommand\thefigure{\arabic{section}.\arabic{figure}}

\maketitle

\begin{abstract}
\singlespacing
\noindent  

Using a state-of-the-art large language model, we extract forward-looking and context-sensitive signals related to inflation and unemployment in the euro area from millions of Reddit submissions and comments. We develop daily indicators that incorporate, in addition to posts, the social interaction among users. Our empirical results show consistent gains in out-of-sample nowcasting accuracy relative to daily newspaper sentiment and financial variables, especially in unusual times such as the (post-)COVID-19 period. We conclude that the application of AI tools to the analysis of social media, specifically Reddit, provides useful signals about inflation and unemployment in Europe at daily frequency and constitutes a useful addition to the toolkit available to economic forecasters and nowcasters. 
\end{abstract}




\section{Introduction\label{sec:introduction}}
Monitoring price and labor market developments is a key task in policy institutions. Since the release of hard data is usually subject to considerable delays, nowcasting models in combination with high-frequency information can be used to bridge the gap. Besides the non-trivial choice of what model to use to generate the nowcasts, practitioners have to select the set of timely information from which they can extract useful signals about the economic status quo. In this paper, we use social network data for nowcasting inflation and unemployment in the euro area.

Our work provides three key contributions: first, we use a large language model (LLM) to process social network data from Reddit to construct forward-looking and context-sensitive indicators related to inflation and unemployment in the euro area. The LLM is notably better at identifying textual information about future economic developments that are relevant for the euro area than other language processing techniques such as dictionaries. Second, we explore  ways to incorporate the role of social interactions in the construction of the time series, using a variety of filters, scoring mechanisms, and decision thresholds to distinguish signals from noise in user-generated comments. We show that accounting for social interaction  refines sentiment analysis and  improves the robustness of economic signals. Third, we rigorously test the out-of-sample validity of the constructed indicators in a nowcasting application. Our findings reveal a consistent out-of-sample improvement in performance with respect to other state-of-the art sentiment indicators and financial variables, especially when we refine our series with social interaction. We achieve gains ranging from 13 percentage points for food price inflation to a minimum of 5 percentage points for youth unemployment.

This paper relates to two strands of research. On the one hand, the nowcasting literature has provided broad evidence that high-frequency, forward-looking variables improve performance in the euro area for important aggregates such as output, inflation, and employment. Studies have used, among others,  surveys \citep{giannone2009nowcasting, banbura2023nowcasting, cascaldi2024back}, newspaper articles \citep{ashwin2024nowcasting, barbaglia2024forecasting}, financial market indicators \citep{breitung2015forecasting, aliaj2023nowcasting}, Google Trends \citep{ferrara2023google}, or price scanner data \citep{beck2024nowcasting}. In spite of the ubiquity of social media and their key role in lowering the cost of information acquisition \citep{han2013social, goyal2017information}, the literature has so far made sparse use of such data for the purpose of macroeconomic monitoring. Notable exceptions are \cite{angelico2022can} who use X/Twitter data for inflation expectations in Italy and \cite{benatti2020high} who leverage LinkedIn data to track European unemployment. Our work adds new insights to this research area by exploring the Reddit social network and evaluating how to use the social interaction arising in the community for signal construction.

On the methodological side, the literature on extracting macroeconomic information from unstructured data has recently started to evolve away from traditional natural language processing (NLP) towards increasingly sophisticated methods such as generative artificial intelligence, for example through the use of LLMs. First-generation NLP methodologies focused on structured extraction and classification of text patterns. Early contributions include \cite{baker2016measuring}, who developed the Economic Policy Uncertainty (EPU) index from newspaper data, and more recent works such as \cite{aprigliano2023power}, who apply topic modeling and frequency analysis to social and news data for real-time indicators, or \cite{barbaglia2024forecasting}, who forecast European GDP with newspaper sentiment extracted with a dedicated economic dictionary. The advent of transformer-based models \citep{vaswani2017attention} has significantly advanced the field by enabling deeper contextual understanding of more general types of documents. Large language models like LLaMa \citep{LLaMa2023} can process and generate high-quality text with minimal or no task-specific fine-tuning, offering a more scalable approach to extracting insights from unstructured textual data. As a consequence, LLMs have found increasing application in the macroeconomic literature as well. \cite{carriero2024macroeconomic} apply LLMs directly to forecast US time series data, while \cite{faria2024artificial} and \cite{hansen2024simulating} show how prompting enables LLMs to act as forecasters, mimicking survey expectations without structured inputs. Closest to our paper, \cite{bybee2023surveying} and \cite{horton2023large} use LLMs to extract sentiment and simulate economic beliefs from US newspaper articles, demonstrating their ability to go beyond pattern recognition and emulate human expectations. We add to this branch of research by applying LLM-driven sentiment analysis for nowcasting macroeconomic variables in the euro area. 

The remainder of the paper is structured as follows: section \ref{sec:methods} explains the raw data extraction and transformation into time series signals using the LLM; section \ref{sec:nowcasting} presents the setup of the nowcasting horse-race and discusses the results in detail; section \ref{sec:conclusion} concludes.

\section{Data and methodology}\label{sec:methods}

\subsection{Collecting social media data}
We use data from Reddit, a social media platform composed of communities, known as ``subreddits", dedicated to specific topics such as politics, technology, and economics. Users can share news items, opinions, and experiences related to economic trends, market behavior, and technological developments, thus producing a rich and timely dataset of user-generated content. A key difference between Redddit and traditional media outlets, is the fact that the users on Reddit make an active selection of the news items/posts they submit. While sentiment indicators that are constructed from newspaper articles reflect the supply side of news generation, Reddit users disclose their interest about selected topics though the selection of posts, thus acting as a potentially important filter of the currently available wealth of information. Moreover, users can engage with the posted content through commenting and voting, which provides an additional layer of signal processing. This combination of characteristics allows us to use Reddit as a source for gauging agents' real-time sentiment.

\begin{figure}[H]
    \centering
    \label{fig_counts}
    \includegraphics[width=17cm, height=10.0cm]{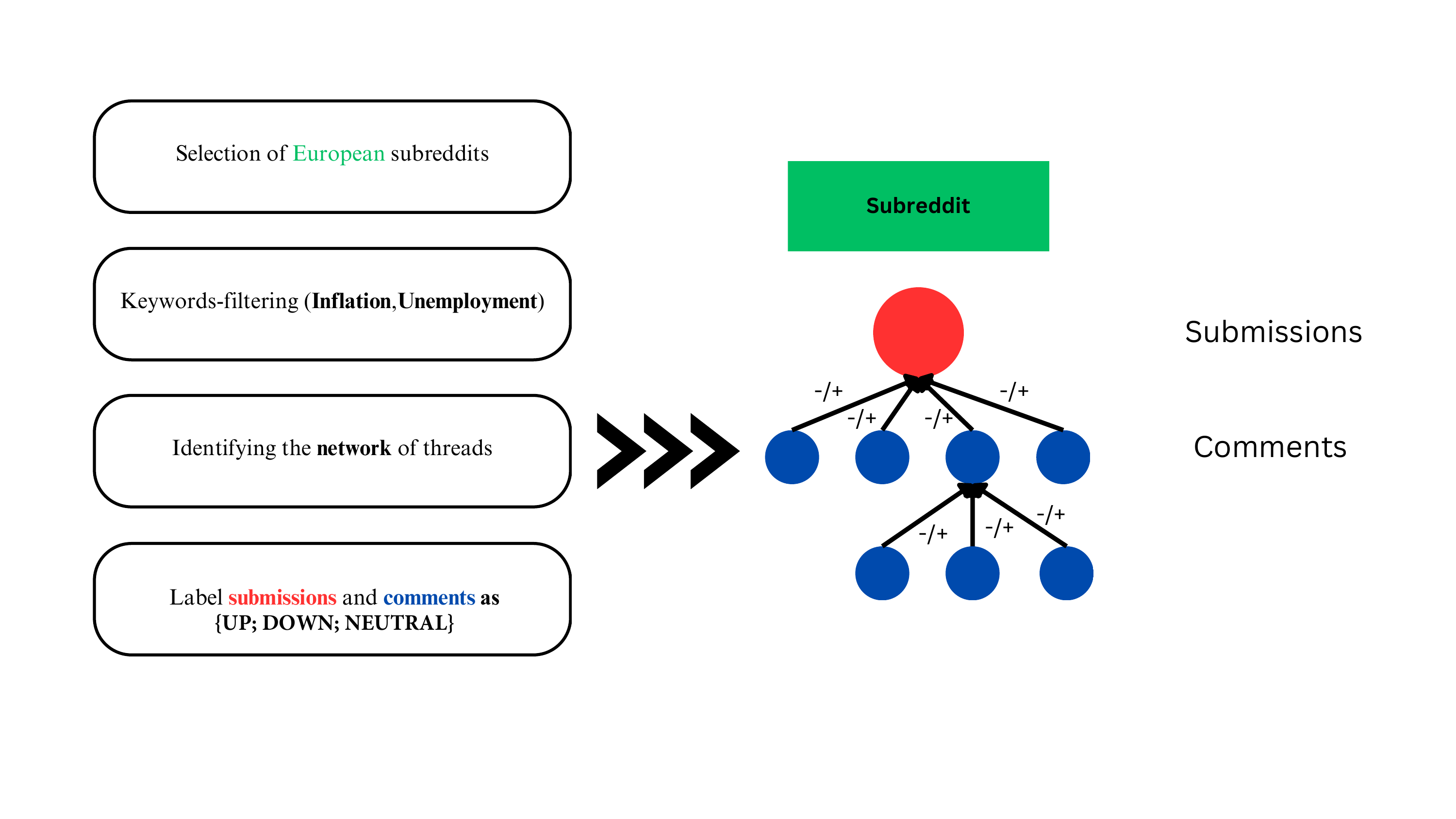}
    \caption{\footnotesize{To extract signals from Reddit data, a four-step approach is employed. First, discussions are sourced from the subreddit \textit{r/europe}. Second, keyword filtering is applied to identify submissions related to predefined economic concepts. Third, comments on each submission are stored. They can be made directly to the submission or to previous comments. We also record the difference between upvotes and downvotes on each comment as a possible weighting device. Fourth, the submissions and comments are then scored as either signaling inflation/unemployment going UP, DOWN, or NEUTRAL by the LLM.
    }}\label{fig_reddit_struct}
\end{figure}

To manage the different levels of information that Reddit has to offer, we proceed in steps, as shown in Figure \ref{fig_reddit_struct}. First, we select the community of interest for our study, the subreddit \textit{r/europe}, which is the largest community discussing European-related news and economic developments. Second, within the selected community, we identify (1) submissions and (2) the comments attached to them.  The first row of Table \ref{tab:summary_subreddits} shows the total number of comments and submissions recorded for the time period between 2012 and 2023. Given the very large number of items, we use a keyword-filtering approach to pre-select the submissions that are relevant for the macroeconomic concepts inflation and unemployment.\footnote{Submissions should contain at least one of the following keywords to be selected: \textbf{inflation-related}: inflation, deflation, hyperinflation, price; \textbf{unemployment-related}: unemployment, employment, unemployed, job. We notice that changing the set of keywords used to filter does not significantly alter the results. Indeed, the most important selecting words turn out to be the economic concept we wish to target, i.e. \textit{inflation} or \textit{unemployment}.} This yields 4,825 inflation-related and 1,934 unemployment-related submissions to be analyzed.

 Third, for the associated comments we make a distinction between those that immediately follow the original submission and the full collection of comments. While the direct comments are likely to be the most relevant ones for the topic raised in the submission, additional layers of comments may still contain valuable information. To make this distinction, we recover the full network structure of comments following an initial submission. We call the unfiltered set of comments that are directly attached to the submission \textit{first-level}. The set of comments across all levels, on the other hand, is very large. Therefore, we make use of the keyword filtering approach discussed above to subset it again. This collection is labeled \textit{keyword}. The final numbers of first level and keyword-filtered comments are summarized in Table \ref{tab:summary_subreddits}. In addition, each comment and submission can be upvoted and downvoted by the users and we record the net scores for all posts as the difference between upvotes and downvotes. Following these steps, the selected submissions and comments are fed to the LLM, to extract a signal that we use to nowcast inflation and unemployment. We explain this step in more detail below.
 
\begin{table}[H]
    \centering
    \resizebox{\textwidth}{!}{
    \begin{tabular}{lccc}
            \multicolumn{4}{c}{\textit{Summary statistics for r/europe}} \\
        \toprule
          & {Submissions} & {Comments} (\textit{first level})  & {Comments} (\textit{keyword}) \\
                  \midrule
                  \textit{r/europe} &  759,179 &  \multicolumn{2}{c}{Total: 25,088,378}   \\ 
        \midrule
        Inflation   & 4,825 & 31,938 & 24,932 \\
        Unemployment  & 1,934 & 14,464 & 12,349 \\    
        \bottomrule
         \end{tabular}
    }
    \caption{Summary statistics for submissions and comments for each concept in the subreddit \textit{r/europe}. The time range is 2012m1-2023m12. }
    \label{tab:summary_subreddits} 
\end{table}

\subsection{Extracting indicators with the LLM}
Due to their generality, LLMs are particularly suited for handling the challenges posed by the linguistic characteristics of Reddit communities, as noted in \cite{long2023just}. Unlike the language found on conventional news outlets, Reddit users often write in community-specific slang, use abbreviations, and generally informal language that strongly deviates from standard dictionaries commonly found in economic research \citep{loughran2011liability}. By using an LLM, we can capture the nuanced, community-specific signals present in Reddit data without the need of extensive fine-tuning or adjustment of existing dictionaries. We use the model \textit{LLaMa-3-70b-instruct} on local servers of the European Commission. For the present research, we instruct the model to classify each submission and comment into one of three categories (UP, DOWN, NEUTRAL) based on the forward-looking signal contained in each of the provided items. The exact prompt is reported below:

\begin{tcolorbox}
\texttt{You are a forecaster and want to predict the \textit{future} $\{concept\}$ in Europe from textual documents. The following document will report sentences potentially referring to Europe: {df[`title'][i]}. You have to print a signal that can be UP (if the document is signalling $\{concept\}$ going up in the short/long-term run), DOWN (if the sentence is signalling $\{concept\}$ going down in the short/long-term run) or NEUTRAL (if the sentence is neutral or does not signal a particular direction on $\{concept\}$). Print only the results of the signal, do not summarize the sentence nor give any reason on your choice. Even if more sentences or paragraphs are provided to you, you only have to print one signal that can be UP, DOWN or NEUTRAL.}
\end{tcolorbox}
\captionsetup{type=figure} 
\captionof{figure}{LLM prompt used with LLaMa-3-70b-instruct. \textit{concept} is either unemployment rate or inflation rate, \texttt{df['title'][i]} contains the submission or comment.}
\setlength{\parindent}{15pt}

As a benchmark, we adopt two dictionary-based approaches to compute the three-outcome signal taking into account positive and negative words appearing in each Reddit submission. In Appendix \ref{app_dictionary} we provide the dictionary, that is based on the work of \cite{granziera2025speaking}. The second dictionary of \cite{consoli2022fine} is substantially more comprehensive and therefore not fully reported in the Appendix. Both works have been shown to be successful in extracting meaningful signals from  economic texts such as Fed speeches \citep{granziera2025speaking} and newspapers \cite{barbaglia2024forecasting}. 

\subsection{Assessing the accuracy of the LLM}
A key benefit of using the LLM over other methods lies in its generality. To quantify the gains over simple dictionary approaches and simultaneously assess the replicability of the result, we calculate the accuracy of the LLM in terms of the F1 score. Since our classification yields three possible outcomes for each piece of text, the F1 score is computed as the unweighted average of the F1 score for each of the three categories. The F1 score per category is computed as
\begin{align}
    \text{Precision } &=\frac{\text{TP}}{\text{TP} + \text{FP}}\\ 
    \text{Recall }&= \frac{\text{TP}}{\text{TP} + \text{FN}}\\
    \text{F1} &= \frac{2\times \text{Precision}\times \text{Recall}}{\text{Precision} + \text{Recall}} 
\end{align}
where TP is the category's true positive rate, FP is the false positive rate and FN is the false negative rate. Evidently, there is no objective ground truth for the sentiment attached to Reddit submissions or comments that could be used to compute and assess error statistics for the LLM. Instead, we have to rely on a human-labeling approach to gauge the accuracy of the models. We randomly select a subset of 483 inflation and 194 unemployment related submissions (10\% of the total number of submissions for the two targets) which we manually label using the UP, DOWN, NEUTRAL scale. We ensure that the time coverage over the selected submissions is even and not bunched in a few specific years. We then compute the F1 scores for the LLM and the two dictionaries explained above. To ensure robustness, we vary the temperature parameter of the LLM on a grid between 0.1 and 0.9 in steps of 0.2. This nonnegative parameter controls the ``creativity" of LLM responses. Higher values -- the maximum being 1 -- typically produce more volatile responses and 0 produces deterministic answers. For each temperature, the LLM is asked to score the same 483/194 submissions 100 times to obtain a measure of variability for each configuration. The dictionary approach of \cite{granziera2025speaking} achieves a score of 0.340 for inflation and 0.333 for unemployment. The \citep{consoli2022fine} dictionary approach reaches 0.290 for inflation and 0.339 for unemployment.

\begin{figure}[H]
    \centering
    \includegraphics[width=\linewidth]{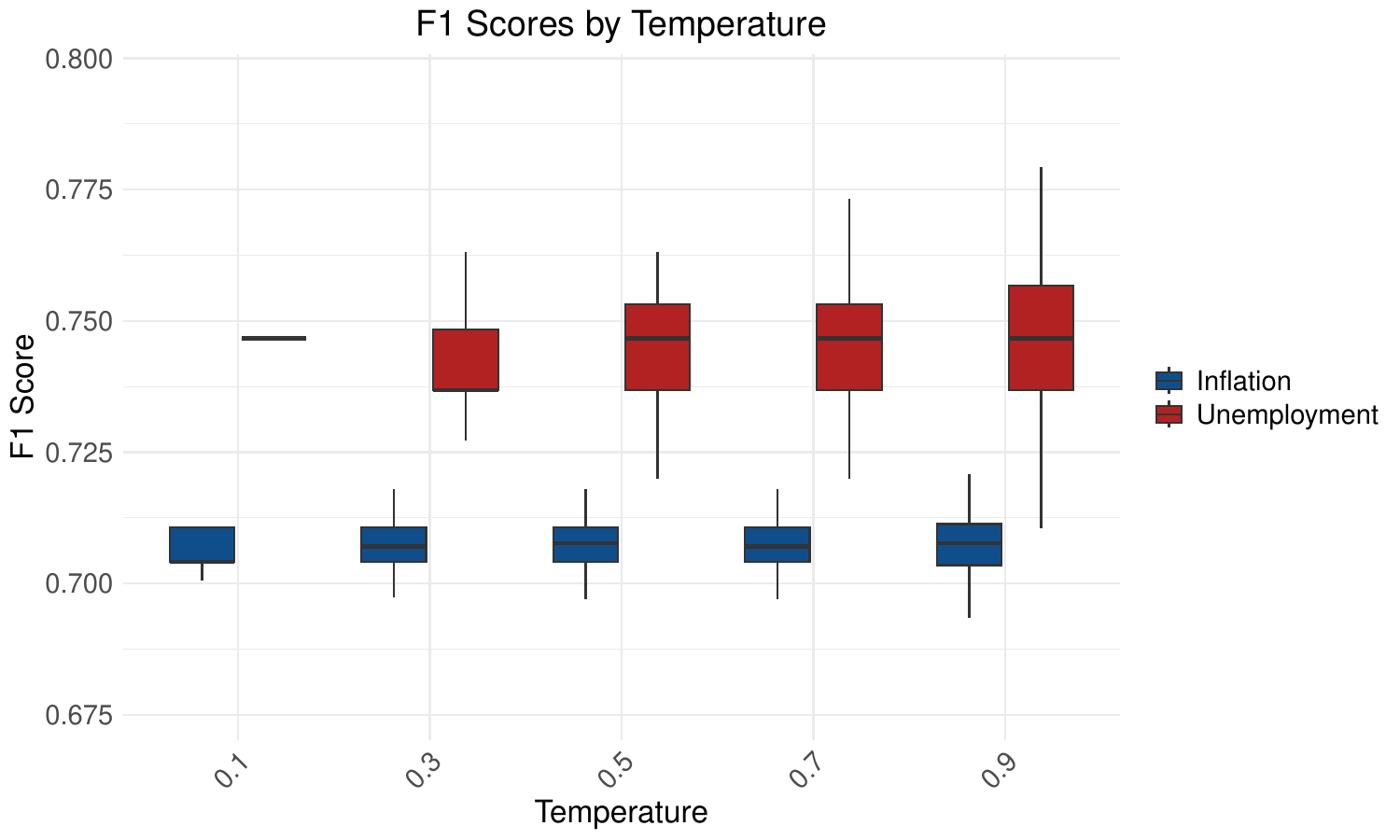}
    \caption{LLM accuracy F1 scores for inflation and unemployment submissions. The model is run 100 times for each temperature value. Boxes represent the interquartile range (Q25 and Q75). Whiskers extend to a 95\% interval for the median (thick horizontal line).}
    \label{fig:accuracy}
\end{figure}

The results for the LLM are visualized in Figure \ref{fig:accuracy}, which shows that it far outperforms the scores of the dictionary approaches for both variables. Unemployment scores are slightly higher than those for inflation. Overall, the scores lie around 0.71 (inflation) and 0.75 (unemployment). In nearly all cases, we can see that increasing the temperature widens the range of possible accuracy scores, while the median stays essentially stable. We have used a temperature of 0.5 throughout the following analysis.

\subsection{Constructing time series indicators} \label{sec:time_series_indicator}
Building on the classifications obtained via the LLM, we construct the time series indicators $X_t$ for submissions only and $\bar{X}_t$ for signals, which combine both submissions and comments. Let us start with the submissions-only signal. To create this, we aggregate all signals $S_i \in \left\{\text{UP, DOWN, NEUTRAL}\right\}$ for submissions posted on day $t$, computing the daily sum, where UP takes value 1, DOWN takes value -1, and NEUTRAL is treated as 0:
\begin{equation}
{X}_{t} = \sum_{i=1}^{N_t} S_i 
\label{eq:llm-indicator}
\end{equation}
$N_{t}$  represents the total number of submissions recorded on a given day. If no submissions are available in $t$, the indicator is assigned a value of zero. $X_t$ is then smoothed using a backward looking moving average (MA) filter of varying window sizes. The effect of the smoothing parameter on nowcasting performance is discussed in Appendix \ref{Appendix:Smoothing}.

\begin{figure}[H]    
    \centering
    \includegraphics[width=16.0cm, height=9.5cm]{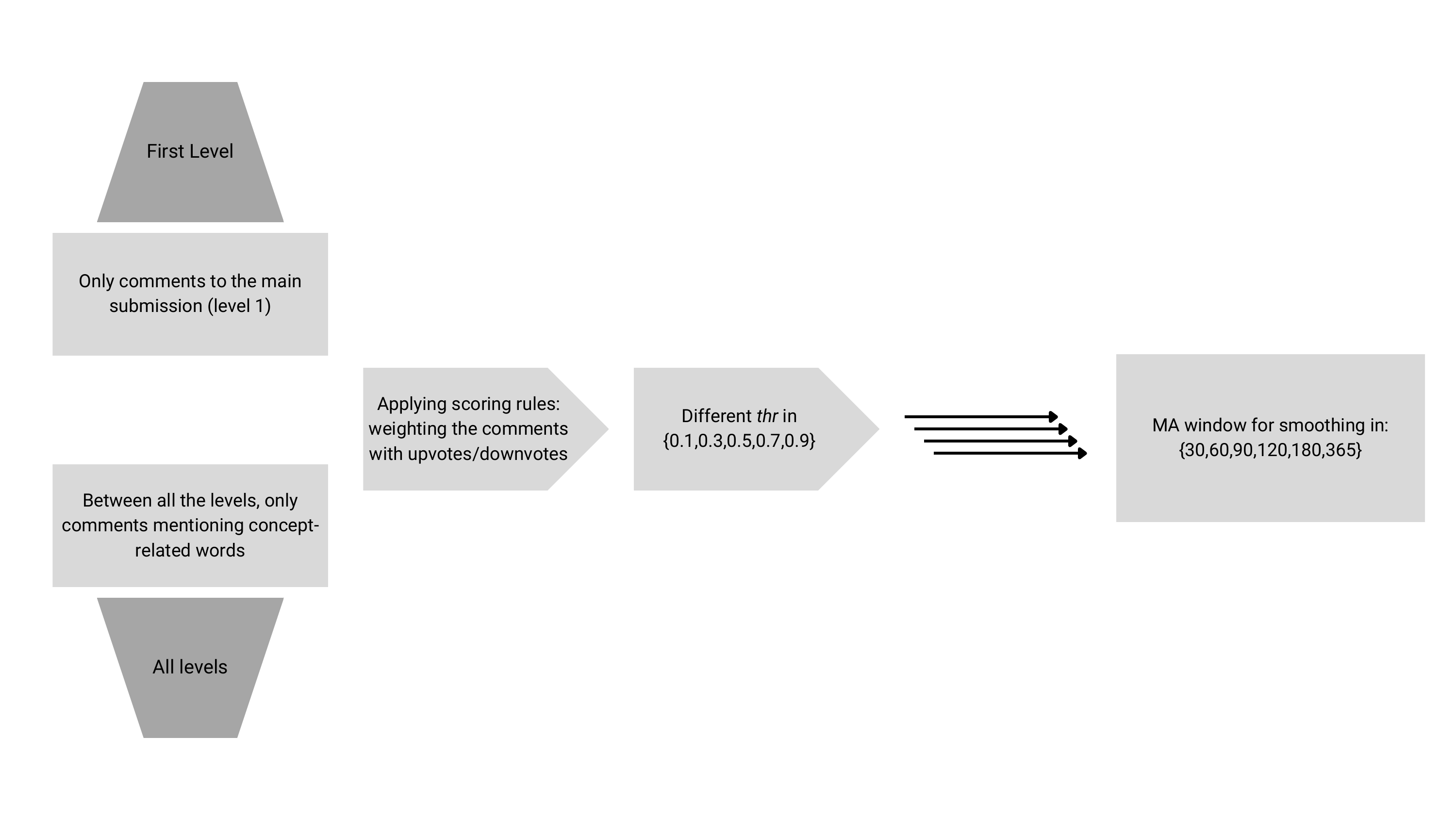}
    \caption{ 2 sets of comments a) all  first level comments; b) considering all the levels, only keeping comments containing keywords related to the concept. To both types of comments the following additional transformation are applied: i) different scoring rules for aggregation based on up- downvotes, ii) different thresholds, iii) transformation with MA smoothing windows.}
    \label{fig_comment_diagram}
\end{figure}

Next, we turn to the construction of time series signals $\bar{X}_t$, which also incorporate social interaction, a key novelty of this paper. The steps are illustrated in Figure \ref{fig_comment_diagram}. 

To do this, for each submission $i$ we compute a score ${L}_{i} \in [-1,1] $, which combines not only the submission's original signal ($S_{i}$), but also the signals derived from the $J$ comments $C_{i,j} \in \left\{\text{UP, DOWN, NEUTRAL}\right\}$ associated to it. If a submission receives no comments, the original signal remains unchanged.\footnote{A valid concern is that comments could arrive much later than the submission, invalidating their use for a nowcasting or forecasting application. We checked the timing of all the comments and found that they are posted at most one week after the main submission. This ensures a significant reduction of any possible look-ahead bias coming from comments arriving at later points in time. To corroborate this, the weekly information flow is discussed in Appendix \ref{Appendix:DailyInforamtion}.}

\begin{equation}\label{eq_aggr_comments}
{L}_{i} = \frac{S_{i} + \sum_{j=1}^{J} C_{i,j}}{J+1}.
\end{equation}

Comments $C_{i,j}$ can either be those that are directly attached to the submission, but unfiltered by keywords (\textit{first level}), or the set of all keyword-filtered comments (\textit{keyword}), as explained in section \ref{sec:methods}. We treat each comment as a vote. If the LLM assigns UP to the comment, the vote is counted as 1. If it assigns DOWN, the vote is -1 and 0 for NEUTRAL classifications. As noted in section \ref{sec:methods}, comments can also receive upvotes and downvotes by users. The net score of these can be used as a weight on the vote that each comment casts. Taking the average of the (possibly weighted) votes of all comments in addition to that of the original submission yields the score ${L}_{i}$ in (\ref{eq_aggr_comments}). Finally, we need a decision rule to transform the score $L_i$ back to the UP, DOWN, NEUTRAL scale. We achieve this through the threshold parameter $\tau$. This parameter is varied on a discrete grid, to evaluate its impact on the quality of the signal. Based on the voting outcome, the new label $\bar{S}_i$ for submission $i$ is obtained:

\begin{equation}
    \bar{S}_i =
\begin{cases}
\text{UP} & \text{if } {L}_{i} > \tau \\
\text{DOWN} & \text{if } {L}_{i} < -\tau \\
\text{NEUTRAL} & \text{if } -\tau \leq {L}_{i} \leq \tau
\end{cases}
\end{equation}

Once this process is finished and all new submission labels $\bar{S}_i$ are obtained, the daily scores are re-computed as $\bar{X}_{t}$ according to:

\begin{equation}
\bar{X}_{t} = \sum_{i=1}^{N_t} \bar{S}_i 
\label{eq:llm-indicator-reclassified}
\end{equation}
The resulting signals $\bar{X}_{t}$ are also smoothed with an MA filter of different window lengths.

\begin{figure}[H]
    \centering
    \begin{subfigure}[b]{0.49\textwidth}
        \centering
        \includegraphics[width=\textwidth]{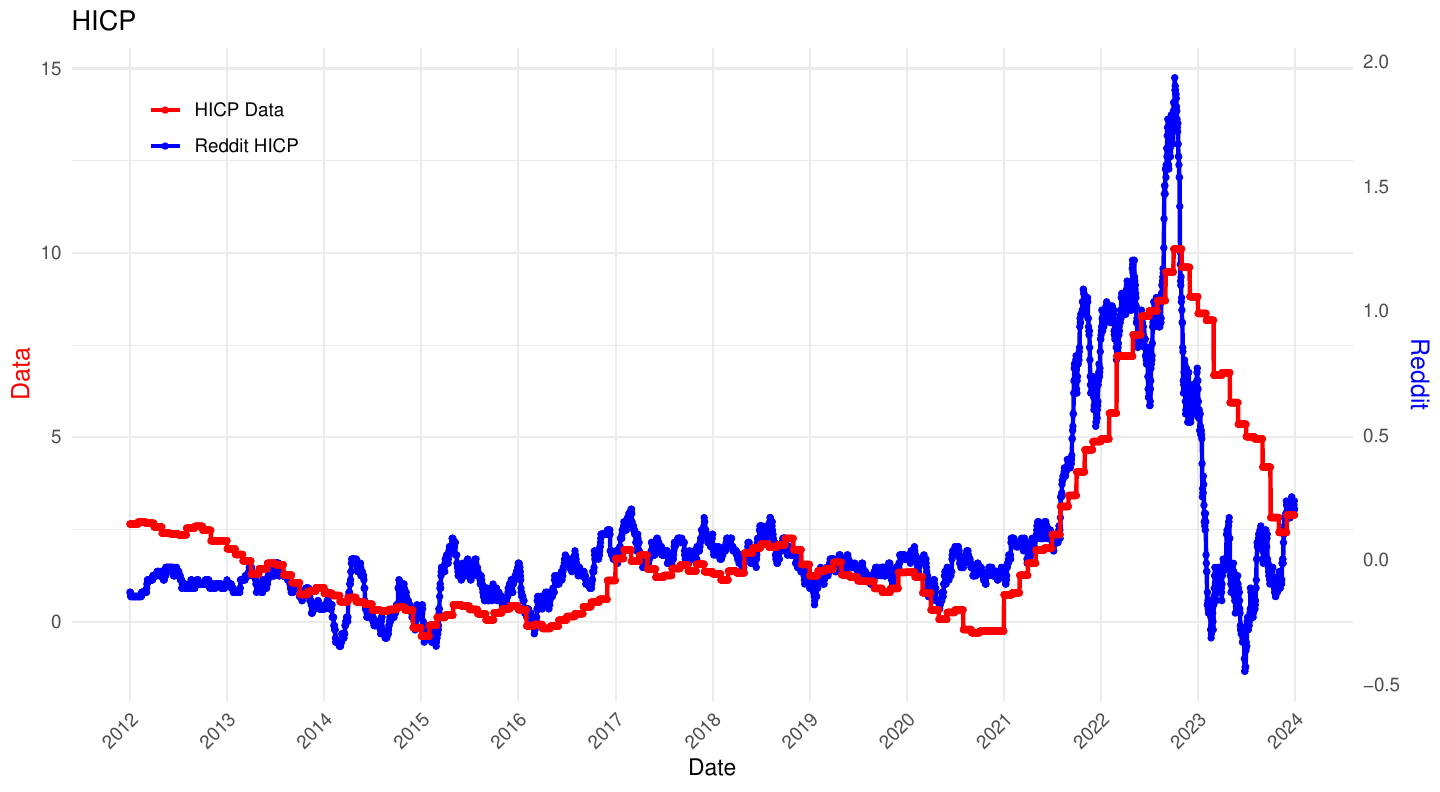}
        \caption{HICP inflation}
        \label{fig:plot1}
    \end{subfigure}
    \hfill
    \begin{subfigure}[b]{0.49\textwidth}
        \centering
        \includegraphics[width=\textwidth]{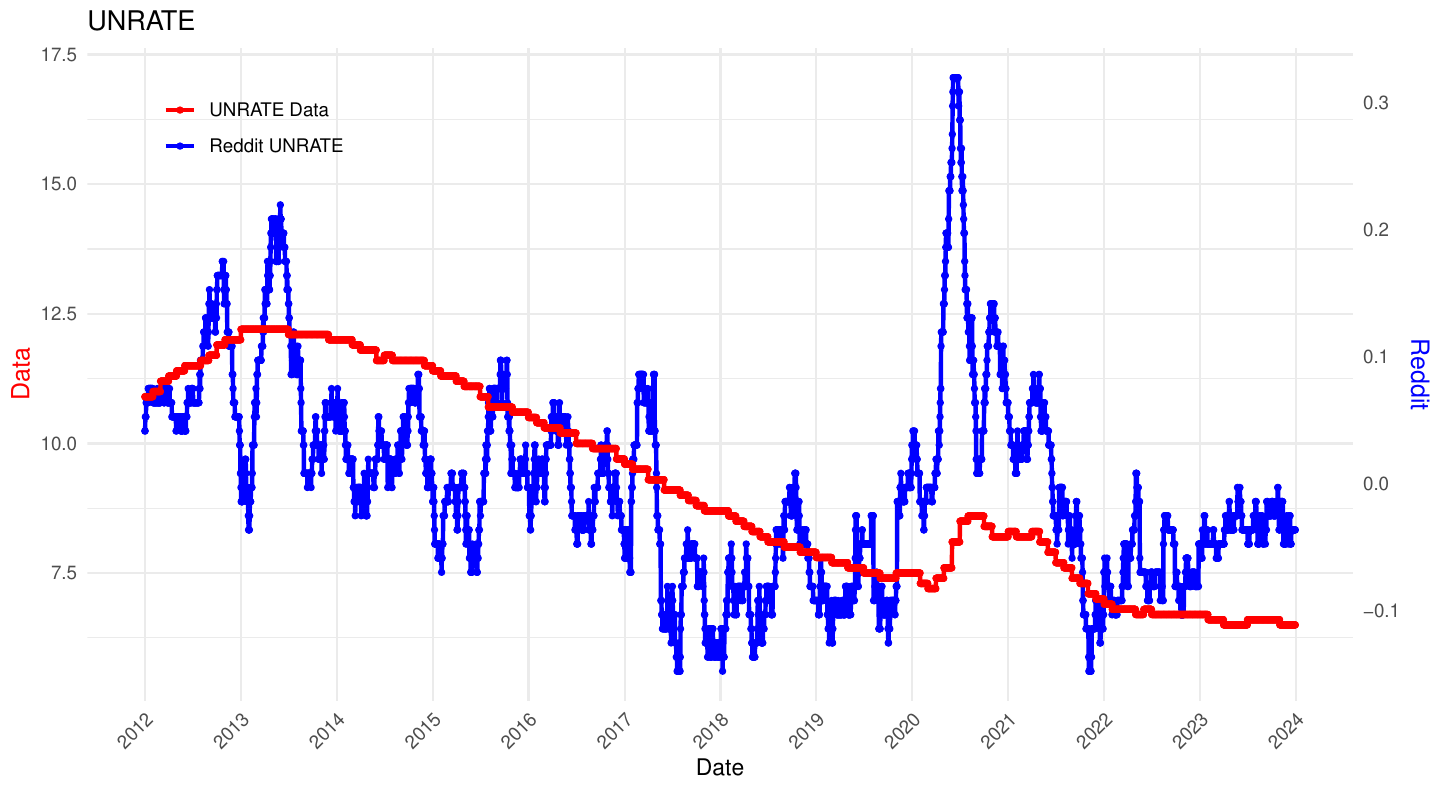}
        \caption{Unemployment Rate}
        \label{fig:plot2}
    \end{subfigure}
    \hfill   
    \caption{Time series of observed HICP inflation and unemployment data at monthly frequency against daily Reddit signals. The Reddit signals are constructed as above and represent the horse-race winners among the 120 specifications.}\label{fig:timeseries}
\end{figure}

Crucially, this voting scheme allows users to overturn the originally assigned score of the submission $S_i$, if they disagree with the original sentiment. Therefore, social interaction acts as regularization device, which introduces checks-and-balances into the signal construction. An example of such a re-classification is presented in Appendix \ref{Appendix:GraphSignal}. In sum, the time series involving not only submissions, but also social interaction through comments differ along four dimensions: (1) unfiltered first level or keyword-filtered comments, (2) usage or not of up- and downvotes, (3) thresholds used for voting outcomes, and (4) MA smoothing windows. All of these choices are evaluated out-of-sample and results are reported in Appendix \ref{Appendix:SocialInteraction}. At the end we have a total of 120 series for each of the two target variables, inflation and unemployment. Examples of the daily signals from Reddit for inflation and unemployment are reported in Figure \ref{fig:timeseries}. Our Reddit signals track HICP inflation and the low frequency component of the unemployment rate remarkably well. Therefore, we proceed by evaluating the informativeness of the signal in an out-of-sample application next.

\section{Nowcasting inflation and unemployment}\label{sec:nowcasting}

\subsection{General setup}
To test whether our Reddit signals are useful for real-time monitoring of macroeconomic conditions in the Euro area, we set up a nowcasting application. The target variables are year-on-year HICP inflation and its components for energy, services, food and core, as well as the overall unemployment rate plus those for people over and under the age of 25 years in the euro area. Target variables are obtained from the EA-MD data base created by \cite{barigozzi2024ea}, with the exception of food price inflation, which comes from the European Commission. These are useful nowcasting targets, as inflation data are released in the middle of the month after the reference period (a flash estimate becomes available around the end of the reference period/the beginning of the following month) and unemployment rate data are typically released one month after the end of the reference period. Moreover, past research has shown that price and labor market developments are among the topics about which people care most deeply (see section IV in \cite{barsky2012information}). Engagement around these topics on social media may therefore plausibly contain early signals of ongoing macroeconomic dynamics. Since Reddit users tend to be relatively young non-experts, we would expect that opinions related to readily observed indicators such as food prices, energy prices or youth unemployment would provide  better results relative to those related to broader inflation and unemployment measures (see also \cite{d2021exposure}). The out-of-sample period ranges from January 2018 to December 2023, leaving an initial estimation sample from January 2012 to December 2017, which we then expand recursively. 

Our set of candidate indicators includes all 120 Reddit indicators whose construction is detailed in section \ref{sec:time_series_indicator}, taken one by one. To this set of social media variables we add for comparison the daily newspaper sentiment indices computed by \cite{barbaglia2024forecasting}, inflation swaps for 1 to 5 years and the Brent oil price (FRED mnemonic DCOILBRENTEU), reflecting the fact that  commodity prices have been shown to aid forecasting performance for inflation \citep{breitung2015forecasting}. The newspaper sentiment indicators of \cite{barbaglia2024forecasting} are available for Germany, France, Italy, and Germany. To obtain an aggregate index per concept, we take the first principal components of the available data. We have considered indicators for inflation, unemployment and monetary policy sentiment.

We use MIDAS regressions to combine monthly data with daily data. In particular, we estimate the following MIDAS-AR regression \citep{ghysels2016mixed}:
\begin{align}
    \alpha(L)y_t = c +  \beta(L) \sum_{i=0}^k w_i x_{s(t)-i} + \epsilon_t, \quad \forall i, \quad w_i=h(\gamma,i) \text{ and } \sum_{i=0}^k w_i = 1
\end{align}
where we choose standard second-degree Almon polynomials to shrink the parameters for the weights:
\begin{align}
    w_i = \frac{\exp\left(\sum_{j=1}^d \gamma_j i^j\right)}{\sum_{l=0}^k \exp\left(\sum_{j=1}^d \gamma_j l^j\right)}, \quad d=2
\end{align}

For each target-predictor combination, the lag order $q$ in $\beta(L)$ for the daily data is determined based on the Akaike information criterion (AIC) each time a new nowcast is produced. The lag order $p$ in $\alpha(L)$ for the endogenous data is fixed at 1. We choose this simple setup because the emphasis of our exercise is on comparing information sets rather than models. Moreover, we note that the exercise is not fully real-time, as we do not use data vintages. However, price indexes and unemployment rates are hardly subject to revisions. For Reddit data (and our other daily indicators) revisions are also negligible and arise only if certain submissions or comments are removed by the moderators or if users delete them themselves. Additionally, as our focus is on assessing the information content of individual Reddit signals and the role social interaction plays, we have not explored the potential benefits of including more than a single Reddit indicator at a time in the nowcast. This could be implemented using variable selection techniques as in \cite{babii2022machine}. We leave this extension for future research. 

The information setting of the nowcasting evaluation is as follows: in a given month, we consider the monthly inflation and unemployment data up to the previous month to be fully observed by the nowcaster. In addition, the nowcaster gets access to daily data through the entire current month. For example, in February 2025, the nowcaster wants to obtain predictions for inflation and unemployment. She has access to monthly observations for both variables through January 2025 and to daily signals all through February 2025. In the case of inflation this reflects the availability of the flash estimate, which comes out between the end of January 2025 and the beginning of February 2025. For the case of unemployment data, our approach marks a conservative benchmark, as we are providing data that is technically released a few weeks later, thus reducing the comparative edge of daily data. 

We evaluate the usefulness of the Reddit signals by comparing the root mean squared forecast errors (RMSFE), the mean absolute forecast errors (MAFE) and the continuously ranked probability score (CRPS) of the MIDAS regressions with those of a nowcast based on a monthly AR(1) model, which serves as a benchmark. For the density forecasts we assume normal errors and simulate 1000 nowcasts per model for each period in the evaluation sample. We assess the difference in nowcasting performances for all three error metrics using the Diebold-Mariano test in the version of \cite{harvey1997testing}. While our models are nested, we rely on the arguments made in \cite{clark2015macroeconomic} that this constitutes a conservative approach to significance testing. Moreover, since the models are nested we use the p-values of one-sided tests in which the AR(1) errors are used as the baseline and the MIDAS nowcast errors as the alternative.

\subsection{Daily horse-race results}

We begin by presenting the results for the eight target variables for the best-performing daily Reddit specification, and we compare them with the best performers among sentiment indicators, financial indicators (swaps), and the oil price.

\begin{table}[H]
\centering
\begin{tabular}{llccccc}
 \multicolumn{6}{c}{\textsc{Nowcasting Results of Daily Horse-Race}} \\
\cmidrule{1-6}
\textit{Target} & \textit{Metric}  & \textit{Best Reddit} & \textit{Best Sentiment} & \textit{Best Swap} & \textit{Oil Price} \\
\midrule
\multirow{4}{*}{\textit{HICP}} &  & \textit{com\_60\_0.3\_0\_1} & \textit{inflation} & \textit{swap1Y} &  \\
 & RMSFE & 0.723$^{**}$ & 0.932 & 0.85$^{*}$ & 0.996 \\
 & MAFE & 0.776$^{**}$ & 0.952 & 0.866 & 1.000 \\
 & CRPS & 0.751$^{**}$ & 0.966 & 0.869 & 1.016$^{**}$ \\
 \midrule
\multirow{4}{*}{\textit{HICP Core}} &  & \textit{com\_365\_0.1\_0\_0} & \textit{inflation} & \textit{swap2Y} &  \\
 & RMSFE & 0.844$^{***}$ & 0.941$^{*}$ & 0.918$^{*}$ & 1.000 \\
 & MAFE & 0.801$^{***}$ & 0.923$^{*}$ & 0.938 & 1.037$^{***}$ \\
 & CRPS & 0.815$^{**}$ & 0.949 & 0.951 & 0.996 \\
 \midrule
\multirow{4}{*}{\textit{HICP Energy}} &  & \textit{com\_60\_0.9\_1\_1} & \textit{monetary policy} & \textit{swap5Y} &  \\
 & RMSFE & 0.847$^{**}$ & 0.997 & 1.028 & 1.011 \\
 & MAFE & 0.814$^{***}$ & 0.998 & 0.982 & 1.003 \\
 & CRPS & 0.832$^{***}$ & 1.014 & 1.011 & 1.006 \\
 \midrule
\multirow{4}{*}{\textit{HICP Food}} &  & \textit{com\_365\_0.7\_1\_1} & \textit{inflation} & \textit{swap1Y} &  \\
 & RMSFE & 0.679$^{***}$ & 0.809$^{**}$ & 0.871$^{**}$ & 1.022$^{**}$ \\
 & MAFE & 0.661$^{***}$ & 0.825$^{**}$ & 0.891$^{*}$ & 0.999 \\
 & CRPS & 0.662$^{***}$ & 0.827$^{**}$ & 0.89$^{*}$ & 0.999 \\
 \midrule
\multirow{4}{*}{\textit{HICP Services}} &  & \textit{com\_365\_0.1\_0\_0} & \textit{inflation} & \textit{swap2Y} &  \\
 & RMSFE & 0.889$^{*}$ & 0.978 & 0.924 & 1.014 \\
 & MAFE & 0.922 & 0.987 & 0.925 & 1.043$^{*}$ \\
 & CRPS & 0.906 & 0.995 & 0.934 & 1.061 \\
 \midrule
 \midrule
\multirow{4}{*}{\textit{Unemployment}} &  & \textit{com\_90\_0.3\_0\_0} & \textit{unemployment} & \textit{swap5Y} &  \\
 & RMSFE & 0.832 & 0.912 & 1.028 & 1.041 \\
 & MAFE & 0.856$^{*}$ & 0.977 & 0.982 & 1.048$^{*}$ \\
 & CRPS & 0.875 & 0.945 & 1.011 & 1.033 \\
 \midrule
\multirow{4}{*}{\parbox{2cm}{\centering \textit{Unemployment}\\$>$ 25}} &  & \textit{com\_90\_0.3\_0\_0} & \textit{unemployment} & \textit{swap1Y} &  \\
 & RMSFE & 0.824 & 0.888 & 1.092$^{**}$ & 1.039 \\
 & MAFE & 0.826$^{**}$ & 0.902$^{*}$ & 1.076$^{*}$ & 1.024 \\
 & CRPS & 0.854 & 0.897$^{*}$ & 1.088$^{**}$ & 0.996 \\
 \midrule
\multirow{4}{*}{\parbox{2cm}{\centering \textit{Unemployment}\\$<$ 25}} &  & \textit{com\_90\_0.3\_0\_0} & \textit{unemployment} & \textit{swap5Y} &  \\
 & RMSFE & 0.906 & 0.962 & 1.040$^{*}$ & 1.034 \\
 & MAFE & 0.872$^{**}$ & 0.957 & 1.016 & 1.000 \\
 & CRPS & 0.909 & 0.974 & 1.060$^{**}$ & 1.000 \\
\bottomrule
\end{tabular}
\caption{Numbers below 1 imply that the benchmark monthly AR(1) is worse than the MIDAS model. Asterisks refer to significance levels 0.10 $(^*)$, 0.05 $(^{**})$, and 0.01 $(^{***})$ as calculated with the one-sided Diebold-Mariano test. ``Best" indicators in each category are chosen on the basis of RMSFE. The best performing indicator per category and per target is in the first row of each target. The structure of the winning specification for the Best Reddit column is ``com\_MA\_threshold\_scoring\_firstlevel'. E.g. ``com\_60\_0.3\_0\_1" means that the best indicator uses comments, not pure submissions, an MA window of 60 days, a threshold of 0.3 to determine the outcome of the comments' inclusion, does not use up/down-votes and considers only the first level of comments, not the full comment structure after the submission.}
\label{tab:horse_race_daily}
\end{table}

Table \ref{tab:horse_race_daily} shows that the Reddit indicator generally outperforms the other variables in terms of point and density accuracy. The nowcasting gains using Reddit range from 13 percentage points (food price inflation) to at least 5 percentage points (youth unemployment) relative to the next best specification. It significantly beats the AR(1) benchmark in most cases and presents particularly strong performance gains for food price inflation and the overall HICP index. Gains for unemployment nowcasting are sizable but statistically significant only for the MAFE. Somewhat surprisingly, the nowcast improvements are slightly larger for the unemployment rate for people \textit{over} the age of 25. Inflation-related newspaper sentiment and swaps also improve frequently over the AR(1), but to a smaller extent than the Reddit indicator. The sentiment index for unemployment proves a solid competitor as well. Models involving oil prices are on par with the AR(1). In all cases, the specification involving not only Reddit submissions, but also comments achieves the best results.

\begin{figure}[H]
    \centering
    \begin{subfigure}[b]{0.99\textwidth}
        \centering
        \includegraphics[width=\textwidth]{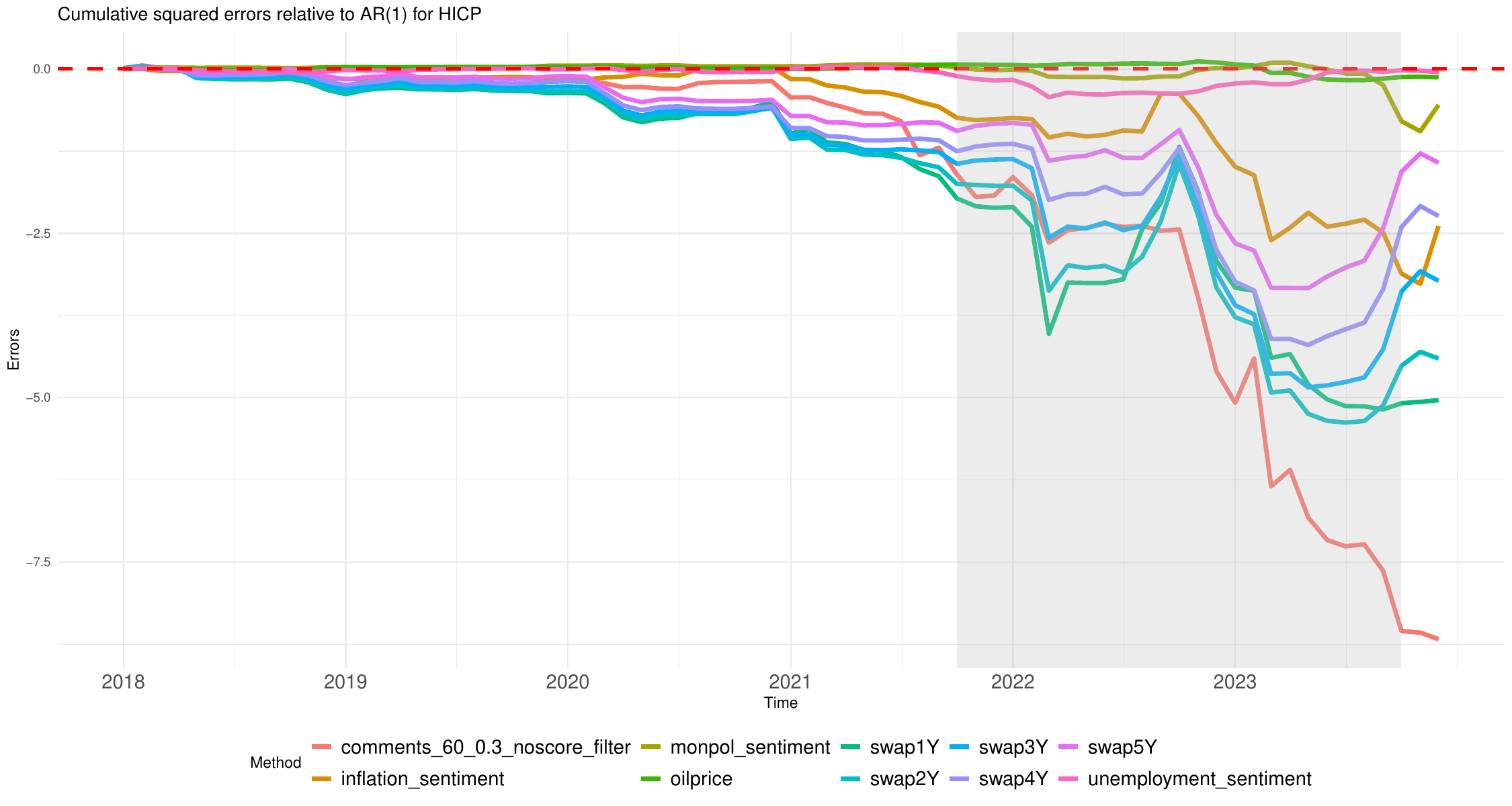}
        \caption{HICP inflation}
        \label{fig:plot2}
    \end{subfigure}
    \hfill
    \vspace{16pt}
    \begin{subfigure}[b]{0.99\textwidth}
        \centering
        \includegraphics[width=\textwidth]{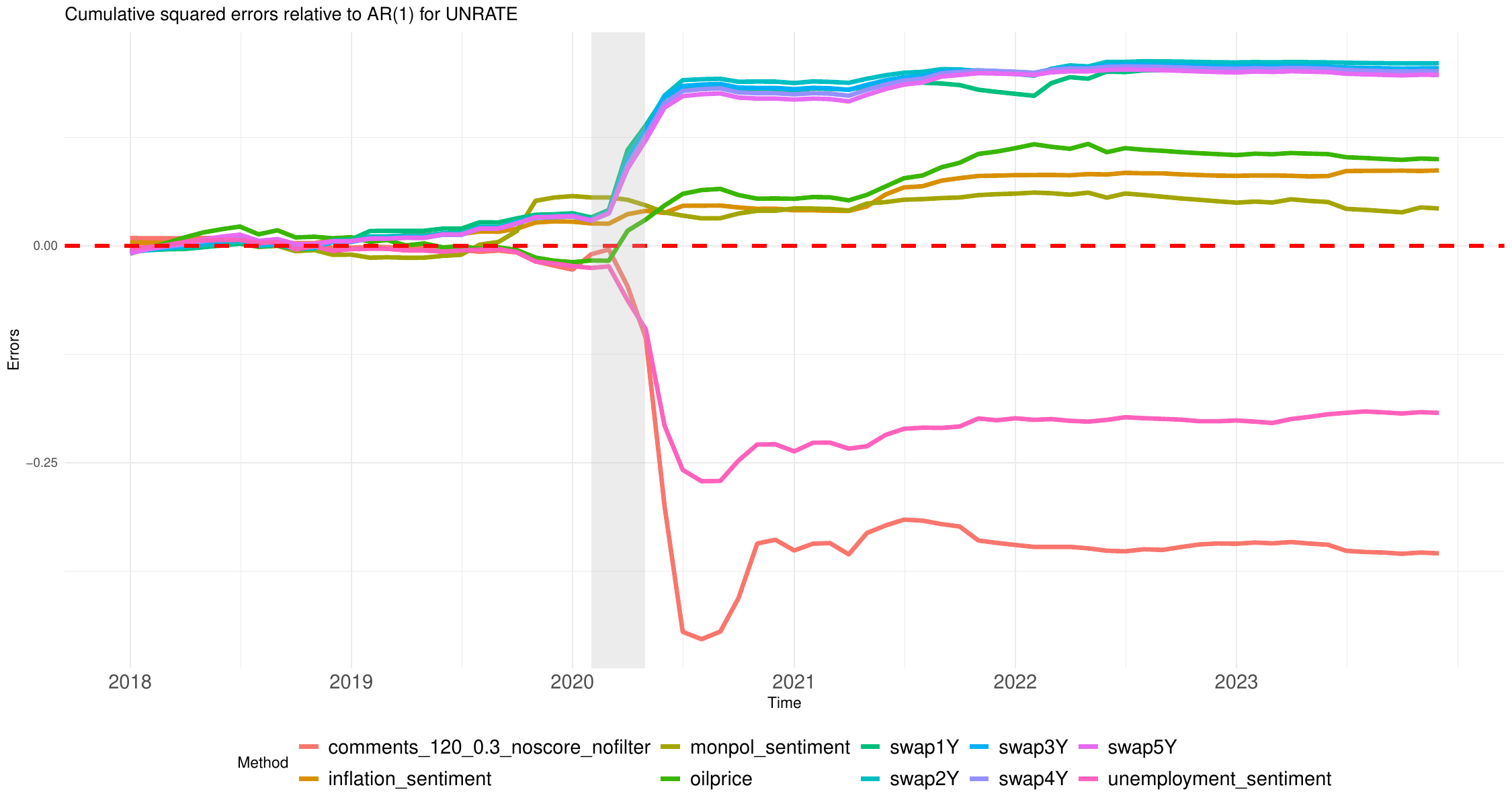}
        \caption{Unemployment rate}
        \label{fig:plot2}
    \end{subfigure}
    \hfill   
    \caption{Cumulative squared errors of MIDAS models minus cumulative squared errors of AR(1) over time. Values below 0 mean better performance than the AR(1). Shaded areas are periods when HICP inflation was above 4\% and the COVID-19 recession months, respectively. Among the Reddit indicators, we provide the cumulative loss for the best performing one according to their target (HICP/Unemployment rate), as shown in Table \ref{tab:horse_race_daily}. }\label{fig:cumulerrors}
\end{figure}

Figure \ref{fig:cumulerrors} shows the evolution of the forecasting performance of the different MIDAS specifications against the AR(1) over time. In terms of HICP inflation, we see that the inflation swaps performed better than the Reddit indicators after the onset of the COVID-19 recession and through the first part of the high inflation period between 2021 and 2023. As the inflation rate starts to drop, this picture reverses and the performance of the Reddit indicator improves markedly. This suggests that the sentiment among social media users reversed faster than on financial markets. For the case of the unemployment rate, we see a single significant improvement in the nowcasting performance after the COVID-19 recession for both the sentiment indicator and the Reddit signal. Otherwise performances are quite stable and the results seem to be driven by a few large shocks that were missed badly by the AR(1). These results appear in line with the literature which has argued that high frequency information is particularly useful in times of turmoil, but not so much in tranquil times \citep{barbaglia2023testing}. We confirm that gains are especially visible during the COVID19 period and the subsequent high inflation episode by means of a Giacomini-Rossi test, reported in Appendix \ref{Appendix:GiacominiRossi}. \\

\subsection{The role of social interaction}
A key take-away from the horse-race is the following: monitoring social media activity around the topics of inflation and unemployment produces a clearer picture of ongoing price and employment dynamics in the euro area. This is particularly true if the data are collected at daily frequency and take into consideration the comments  of other users. In this section we aim to shed some additional light on the way social interaction improves the social media signal. When we compare the classification of all posts based on the LLM using only the original piece of information (submission) and the revised classification which takes into account comments, we observe that in the case of inflation, the submission classification is re-classified in 11\% of the cases. For unemployment the share increases to  15\%. In both cases, social interaction leads to hundreds of re-classifications. Since nothing else changes between the MIDAS model that uses only submissions and the one that accounts for social interaction, it is this re-classification which reduces nowcasting errors.

\begin{table}[ht]
\centering
\begin{tabular}{ccc}
\multicolumn{3}{c}{\textit{Re-classification of Reddit signals for inflation}}\\
\toprule
Submissions & Comments & Number of re-classifications \\
\midrule
-1 & 0  & 154 \\
-1 & 1  & 8 \\
0  & -1 & 43 \\
0  & 1  & 196 \\
1  & -1 & 2 \\
1  & 0  & 124 \\
\bottomrule
\end{tabular}
\caption{Re-classifications of inflation signals attached to a post based (1) only on the original submission or (2) factoring in the sentiment attached to the resulting comments. There are in total 527 re-classifications out of 4825 (10.92\%). Comments uses threshold 0.3, no upvote/downvote scoring and only the first level of comments as this was the winning configuration.}
\label{tab:reclassification_hicp}
\end{table}

\begin{table}[ht]
\centering
\begin{tabular}{ccc}
\multicolumn{3}{c}{\textit{Re-classification of Reddit signals for unemployment}}\\
\toprule
Submissions & Comments & Number of re-classifications \\
\midrule
-1 & 0  & 111 \\
-1 & 1  & 2 \\
0  & -1 & 58 \\
0  & 1  & 73 \\
1  & -1 & 0 \\
1  & 0  & 47 \\
\bottomrule
\end{tabular}
\caption{Re-classifications of unemployment signals attached to a post based (1) only on the original submission or (2) factoring in the sentiment attached to the resulting comments. There are in total 291 re-classifications out of 1934 (15.05\%). Comments uses threshold 0.3, no upvote/downvote scoring and a dictionary filter for comments as this was the winning configuration.}
\label{tab:reclassification_unr}
\end{table}

\begin{figure}[H]
    \centering
    \begin{subfigure}[b]{0.8\textwidth}
        \centering
        \includegraphics[width=\textwidth]{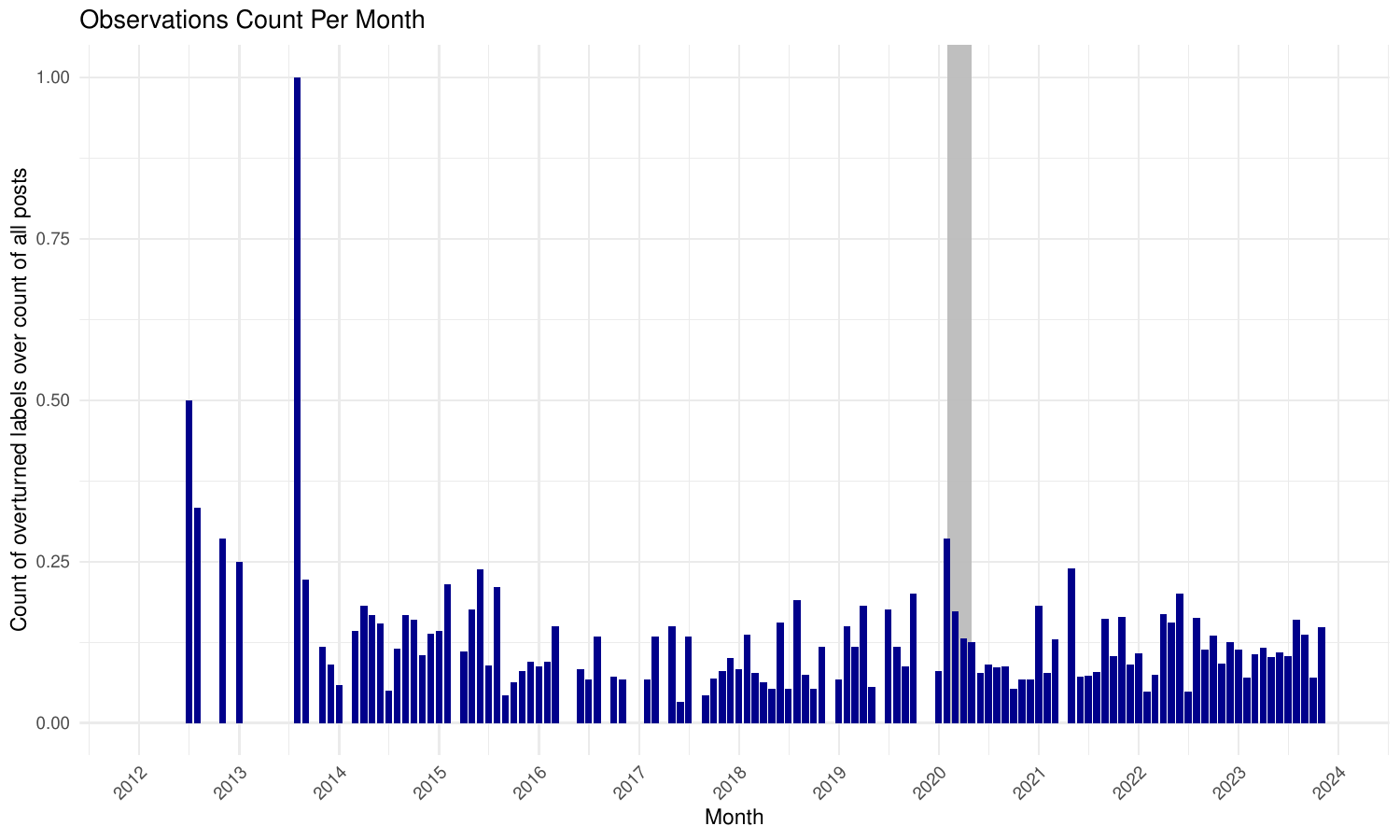}
        \caption{HICP inflation}
        \label{fig:plot1}
    \end{subfigure}
    \hfill    
    \begin{subfigure}[b]{0.8\textwidth}
        \centering
        \includegraphics[width=\textwidth]{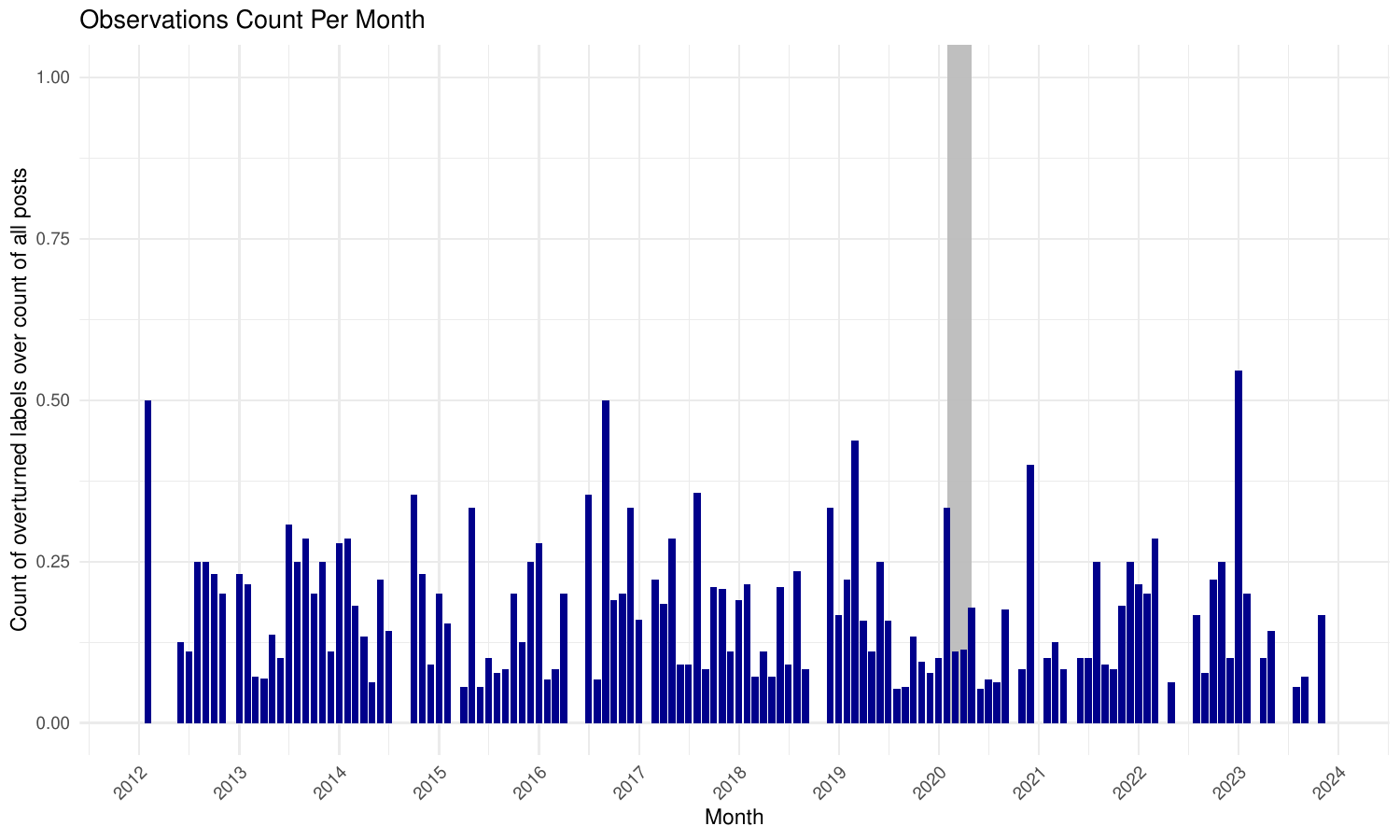}
        \caption{Unemployment rate}
        \label{fig:plot2}
    \end{subfigure}
    \hfill   
    \caption{Frequency of re-classifications as a fraction of total submissions over time. Grey shaded area is the COVID-19 recession.}\label{fig:reclassifications}
\end{figure}

Tables \ref{tab:reclassification_hicp} and \ref{tab:reclassification_unr} show the direction of change that comes about through the incorporation of comments into the signal for the winning specifications. In the case of inflation, the total signal is revised upwards whenever the re-classification is from -1 to 0, -1 to 1 and from 0 to 1. This happens 366 times (counting -1 to 1 twice) while only 171 downgrades occur, a ratio of 2.1. Therefore, the overall sentiment for the direction of inflation in the sample we have considered was generally revised upwards. For unemployment, there are 188 upward revisions and 105 downward revisions, a ratio of 1.8. The results of the horse race suggest that this sharpens the signal, as the nowcast accuracy improves. Since we have observed the clearest nowcasting gains in unusual time periods, we also check if these re-classifications are concentrated in and around these times. Figure \ref{fig:reclassifications} shows that this is not the case, which leads us to conclude that this type of regularization is not episodic but constitutes a systematic feature of social media signals that researchers can use to improve their information set. 

\section{Conclusion}\label{sec:conclusion}
This paper contributes three novelties to the literature. First, by employing a state-of-the-art LLM, we extract forward-looking and context-sensitive signals related to inflation and unemployment in the euro area from millions of Reddit submissions and comments. Relying on this specific data source represents a valuable filter, because users make an active selection of the posts they deem important enough for submission to the network. Moreover, we show that in the context of analyzing diverse economic topics on social media, using an LLM is more accurate than other popular language processing approaches such as dictionaries. 

Second, we suggest a method to include social interaction among users in the construction of the signals by means of a voting scheme. The sentiment conveyed in the discussions triggered by the initial submissions to the \textit{r/europe} thread can be used to refine the sentiment scores if users disagree with the message about future economic developments contained in the submission alone. We explore different facets of constructing these signals, based on the depth of the comment network, weighting through up- and downvotes, and decision thresholds.

Third, we perform a rigorous out-of-sample evaluation of the signal quality of all constructed social media indicators. Our empirical results show consistent gains in out-of-sample nowcasting accuracy when using the signals which incorporate social interaction, even when compared to daily newspaper sentiment and financial variables. Improvements range from up to 13 percentage points for food price inflation to at least 5 percentage points for youth unemployment, highlighting the added predictive value of social media-derived insights.

We conclude that the application of AI tools to the analysis of the vast, multi-faceted information contained in social media, specifically Reddit, constitutes a useful addition to the toolkit available to economic forecasters and nowcasters. We view as promising areas for future research the use of more sophisticated, possibly multivariate nowcasting models, which allow for the inclusion of more than one Reddit indicator and competitor variables such as Bayesian VARs, Dynamic Factor Models or variable selection methods. Moreover, an extension of the analysis to sub-European Reddit threads may be beneficial for obtaining timely information on economic developments tailored to specific countries.

\newpage

\bibliographystyle{apalike}

\newpage

\appendix
\section*{Online appendix\label{sec:appendix}}
\addcontentsline{toc}{section}{Appendices}
\renewcommand{\thesubsection}{\Alph{subsection}}
\renewcommand{\thetable}{\Alph{subsection}.\arabic{table}}  
\renewcommand{\thefigure}{\Alph{subsection}.\arabic{figure}}  

\subsection{Details on social media signal construction}\label{Appendix:Signal}
\subsubsection{Dictionary-based signal}\label{app_dictionary}

In Table \ref{tab:sentiment_words} we provide the words that reflect positive or negative sentiment/signal towards economic concepts. The dictionary is taken from \cite{granziera2025speaking}.

\begin{table}[H]
    \centering
    \resizebox{0.66\textwidth}{!}{%
    \begin{tabular}{p{7cm} p{7cm}}
        \toprule
        \textbf{Positive Sentiment (+1)} & \textbf{Negative Sentiment (-1)} \\
        \midrule
        Boost & Below \\
        Climb & Collapse \\
        Elevate & Damp \\
        Escalate & Deteriorate \\
        Expand & Decline \\
        Foster & Diminish \\
        High & Down \\
        Increase & Drop \\
        Height & Ease \\
        Intensify & Fall \\
        Jump & Low \\
        Persist & Modest \\
        Pressure & Moderated \\
        Moderate & Muted \\
        Rise & Plummet \\
        Risk & Reduction \\
        Remain & Restrain \\
        Rising & Retreat \\
        Rose & Set Back \\
        Risen & Slow \\
        Soar & Soft \\
        Solid & Subdued \\
        Spike & Weak \\
        Sustain & \\
        Strong & \\
        Strength & \\
        Surge & \\
        Upward & \\
        Up & \\
        Upside Risk & \\
        \bottomrule
    \end{tabular}%
    }
    \caption{Sentiment Classification of Words}
    \label{tab:sentiment_words}
\end{table}

\subsubsection{Signal construction with social interaction}\label{Appendix:GraphSignal}
\begin{figure}[H]
    \centering
    \includegraphics[width=17.4cm, height=10.5cm]{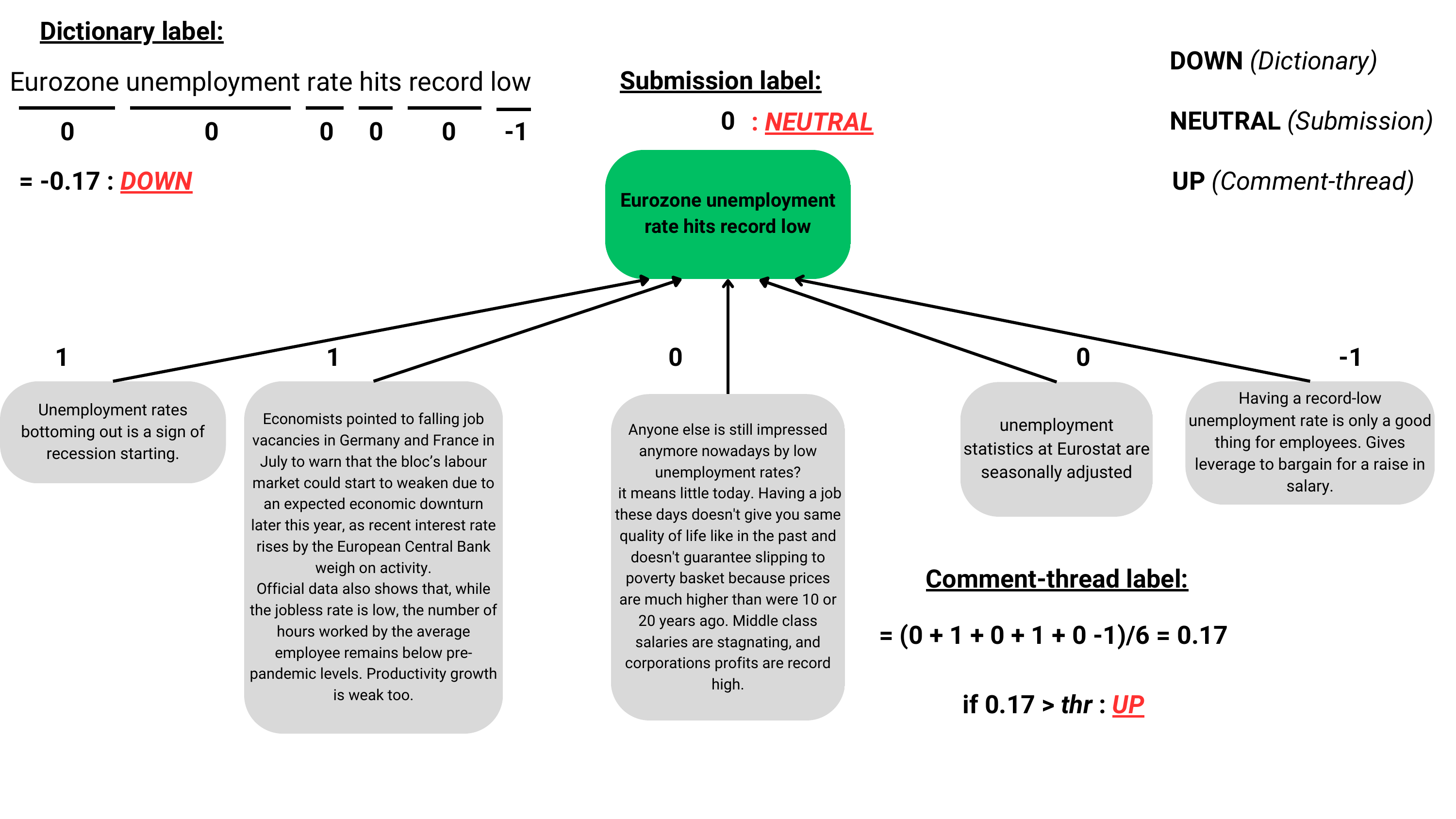}
        \caption{Illustration of a submission-comment network. The initial submission is labeled as {DOWN} (-1) by the dictionary classifier, and {NEUTRAL} (0) by the LLM. The submission's signal is adjusted to UP using comment-based signals and a threshold $\tau = 0.1$.}
        \label{fig_subm_struct}
\end{figure}

To further illustrate the steps involved in the construction of Reddit signals with social interaction, Figure \ref{fig_subm_struct} highlights two points. First, it provides an example of how the LLM classifies a submission differently from a dictionary approach (in this example NEUTRAL (LLM) vs. DOWN (dictionary)). Since we specifically instruct the LLM to predict the \textit{future} direction of the economic concept, it is arguably correct in classifying as NEUTRAL the submission's statement, which is simply a description of the current state of the unemployment rate. The dictionary approach of \cite{granziera2025speaking}, on the other hand, attributes a downward direction to the submission, due to the presence of the keyword ``low". Second, the graphic shows a situation where the sentiment conveyed in the unweighted first-level comments following the submission overturns the score assigned to it. The commenters view the record low of the unemployment rate as a turning point in the business cycle and predict that it will rebound in the future, leading to an overall upward expectation. For a threshold of 0.1, this is sufficiently strong to re-classify the label of the submission from NEUTRAL to UP.

\subsubsection{Signal comparison with other forward looking indicators}\label{Appendix:CompareSentimentSignals}
\begin{figure}[H]
    \centering
    \begin{subfigure}[b]{0.49\textwidth}
        \centering
        \includegraphics[width=\textwidth]{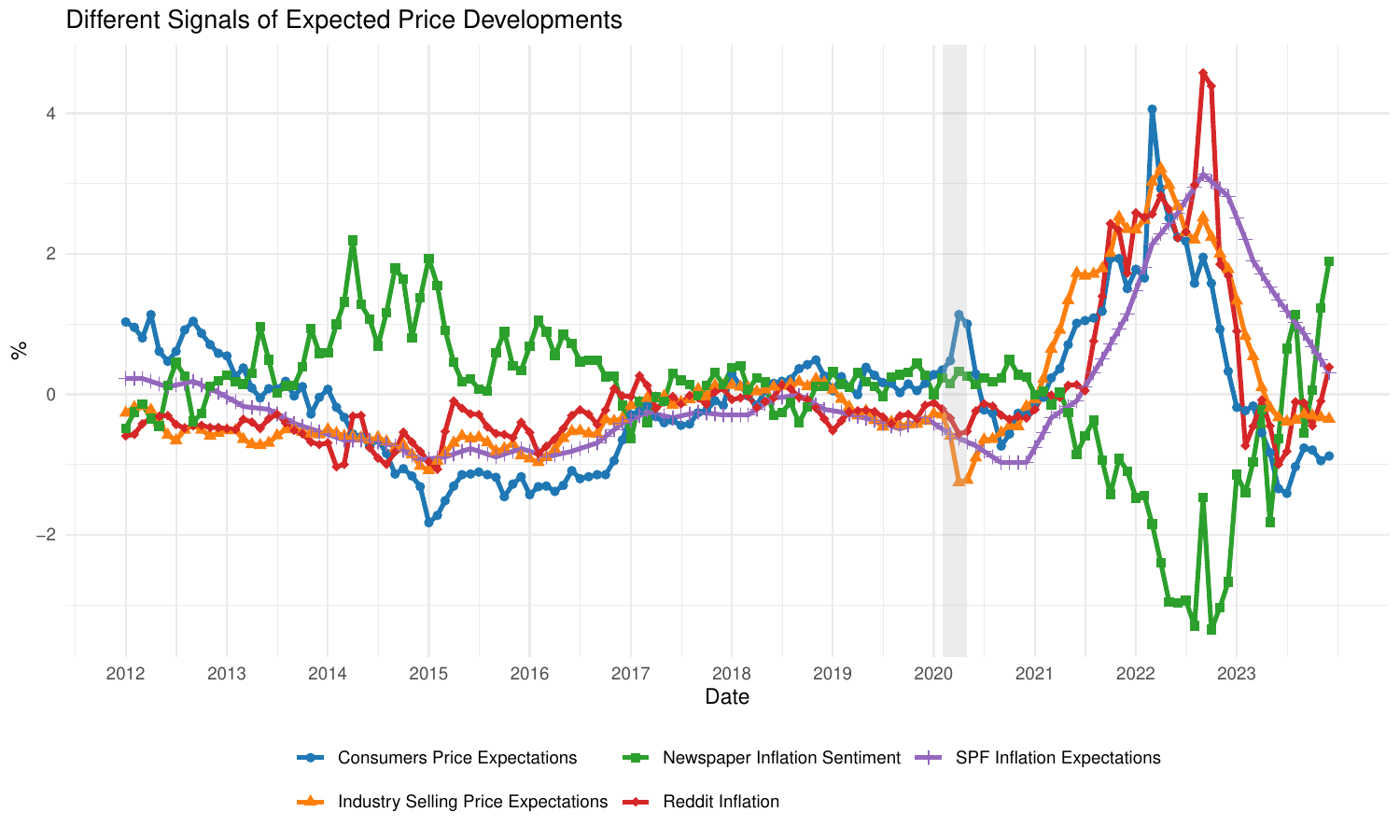}
        \caption{Inflation Expectations}
        \label{fig:plot1}
    \end{subfigure}
    \hfill
    \begin{subfigure}[b]{0.49\textwidth}
        \centering
        \includegraphics[width=\textwidth]{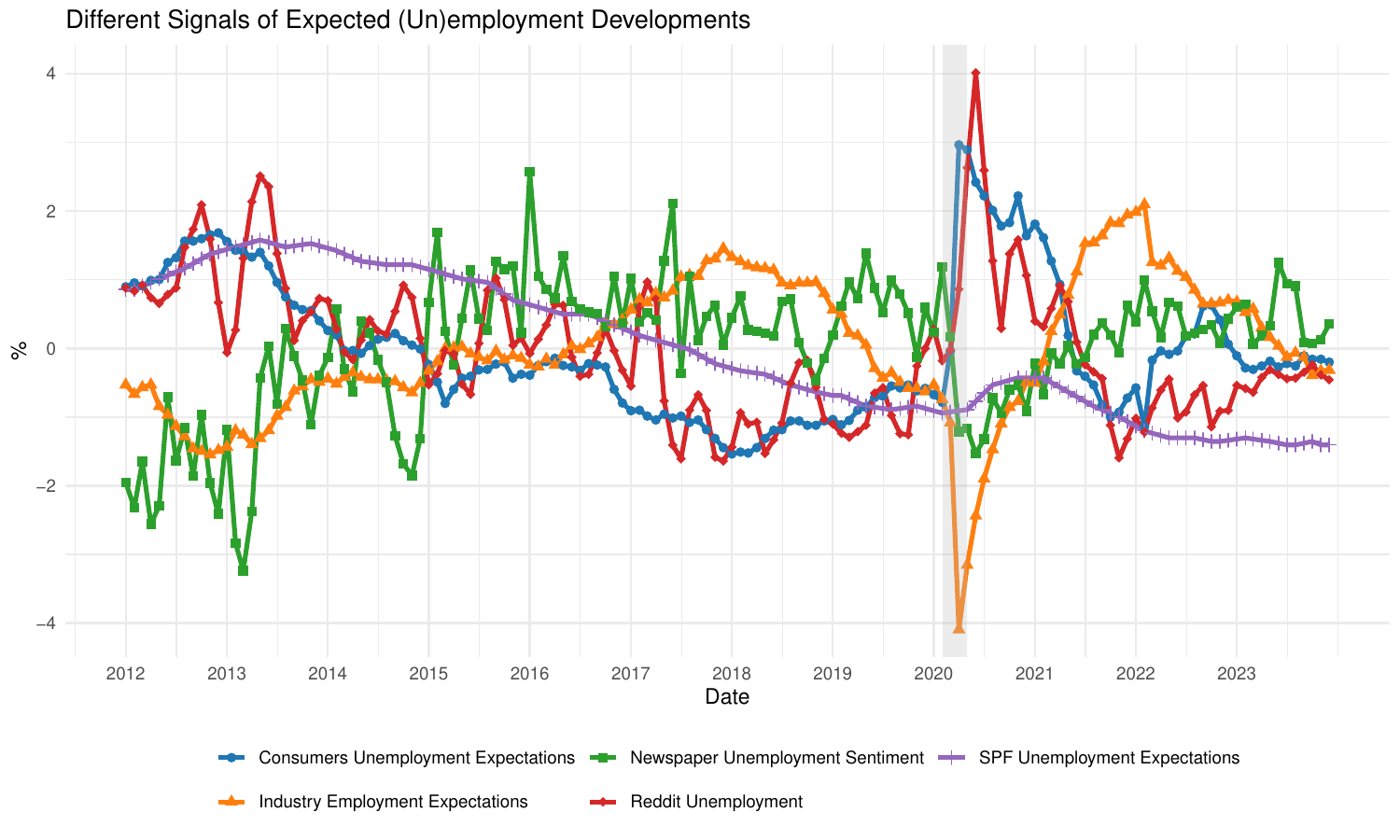}
        \caption{Unemployment Expectations}
        \label{fig:plot2}
    \end{subfigure}
    \hfill   
    \caption{Monthly time series of different measures of inflation and unemployment expectations. Daily data are aggregated to monthly frequency using averages. Consumer and industry data are EU wide totals and come from the business and consumer survey database of the European Commission, SPF data come from the ECB, Sentiment indicators come from \cite{barbaglia2024forecasting} and Reddit series are the authors' own calculations. All series are demeaned and standardized. SPF data is for the reference period ending at the time the survey is conducted and linearly interpolated from quarterly to monthly frequency. Sentiment indicators are the first principal components of the series for Germany, France, Italy, and Spain of the respective category. Shaded areas correspond to the COVID-19 recession. }
    \label{fig:signals}
\end{figure}

A comparison of the (horse-race winning) Reddit series with existing measures of sentiment and expectations, is shown in Figure \ref{fig:signals} and Table \ref{tab:correlations}. Both compare different survey and high-frequency measures of inflation and unemployment expectations with the social media indicators we construct. The Reddit series present high, but not perfect correlations with the other indicators, suggesting that they may convey additional information  that is not present in the other variables. 

\begin{table}[!ht]
    \centering
    \begin{tabular}{l c c}
        \multicolumn{3}{c}{\textit{Correlations of Expectations with Reddit Signals}} \\ \midrule
        & \textit{Inflation} & \textit{Unemployment} \\
        \midrule
        {Consumers Price Expectations} & 0.694 & 0.774 \\
        {Industry Selling Price Expectations} & 0.886 & -0.738 \\
        {SPF Inflation Expectations} & 0.763 & 0.473 \\
        {Newspaper Inflation Sentiment} & -0.783 & -0.562 \\
        \bottomrule
    \end{tabular}
    \caption{Contemporaneous correlation coefficients are rounded to three decimals. Construction of the indices as described in Figure \ref{fig:signals}. }
        \label{tab:correlations}
\end{table}

\subsection{Details on nowcasting horse-race}\label{Appendix:HorseRace}

\subsubsection{Horse race target variables}\label{Appendix:TargetData}
\begin{figure}[H]
    \centering
    \begin{subfigure}[b]{0.49\textwidth}
        \centering
        \includegraphics[width=\textwidth]{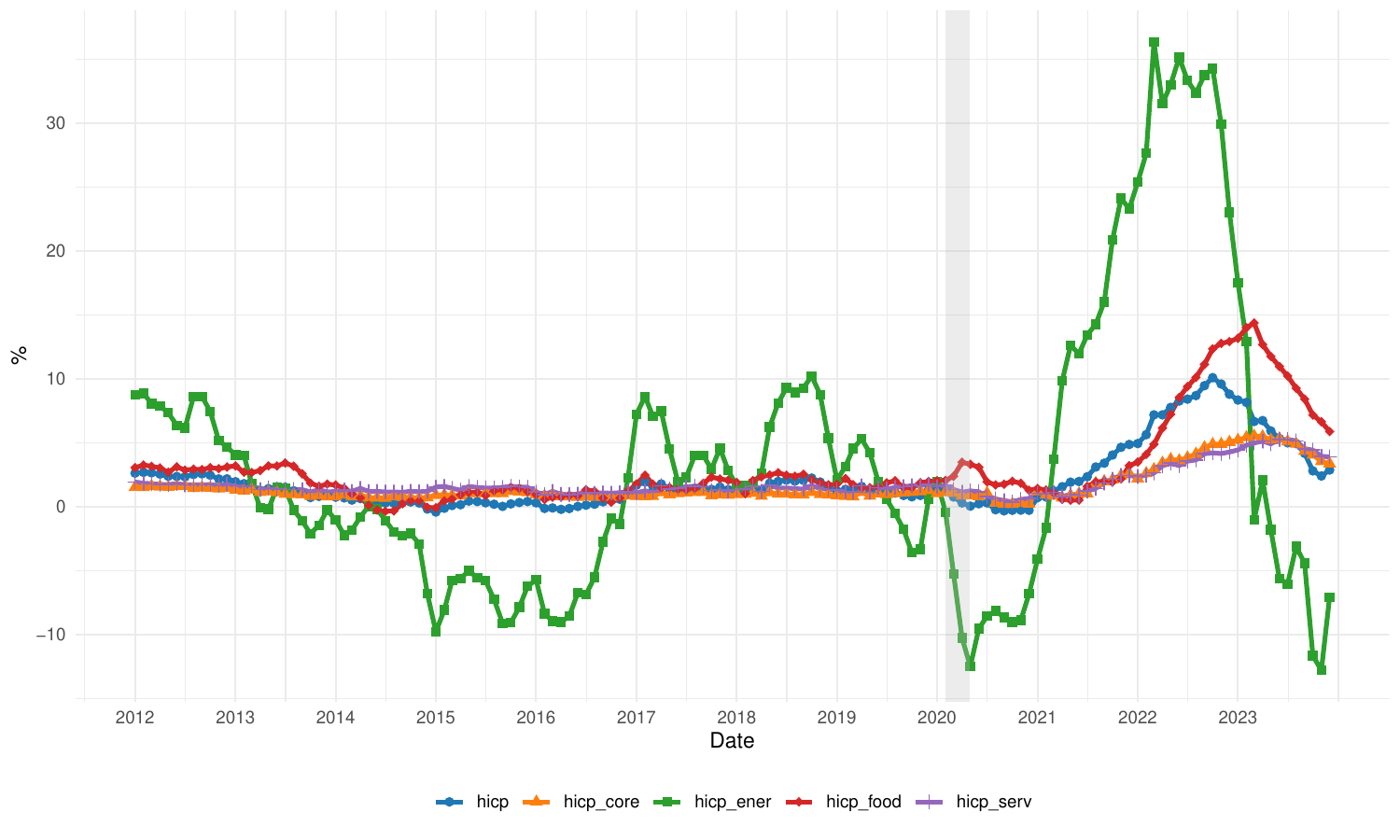}
        \caption{Inflation data}
        \label{fig:plot1}
    \end{subfigure}
    \hfill
    \begin{subfigure}[b]{0.49\textwidth}
        \centering
        \includegraphics[width=\textwidth]{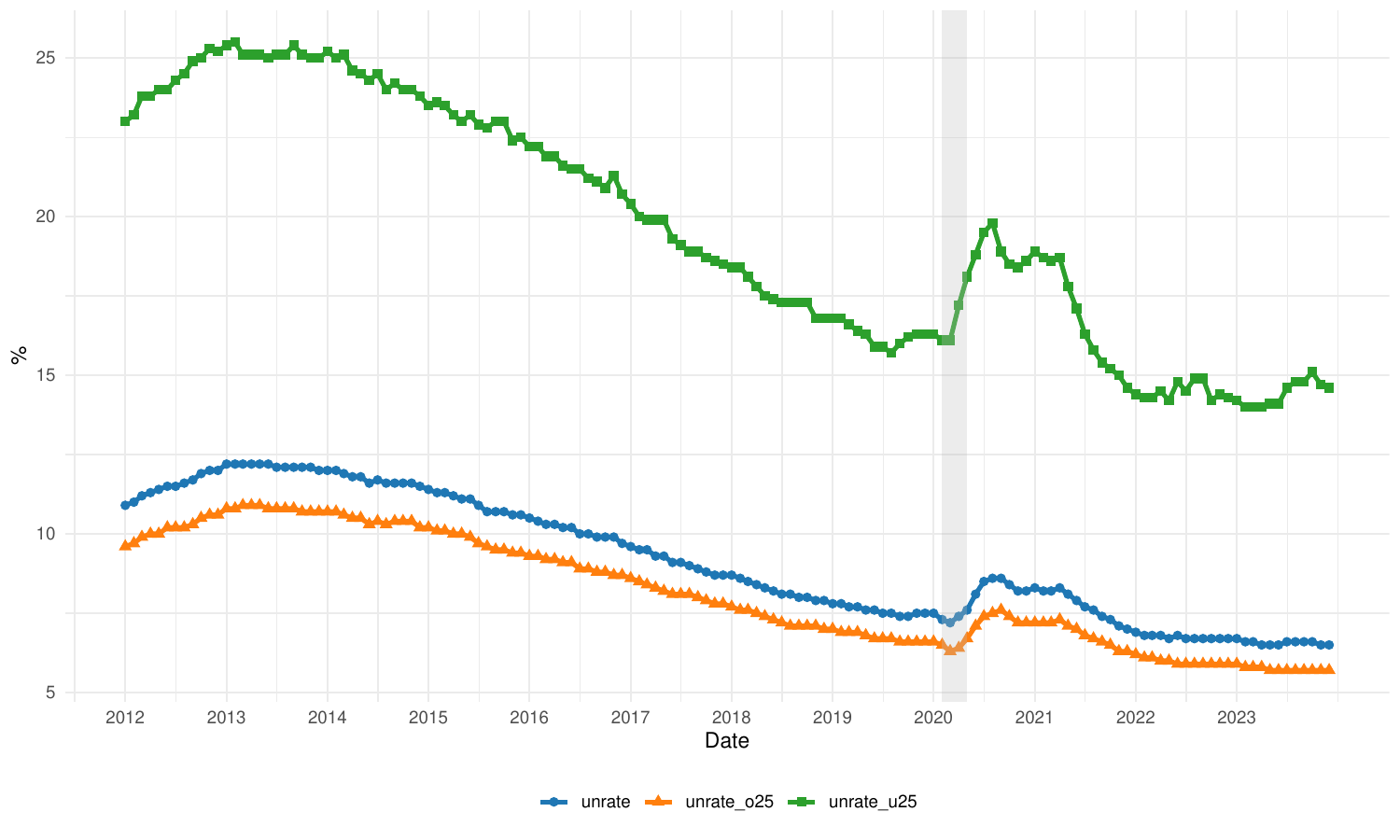}
        \caption{Unemployment data}
        \label{fig:plot2}
    \end{subfigure}
    \hfill   
    \caption{Time series of observed HICP inflation and unemployment data from the EA-MD data base of \cite{barigozzi2024ea}. Grey shaded area is the COVID-19 recession.}\label{fig:series}
\end{figure}

The target variables for the horse-race are depicted in Figure \ref{fig:series}. Clearly energy price inflation was more volatile than the other series in the sample we examine. Similarly, the youth unemployment rate was at a substantially higher level than the aggregate or over 25 years unemployment rates.

\subsubsection{More daily horse-race results}\label{Appendix:MoreHorseRaceResults}
\begin{figure}[H]
    \centering
    \begin{subfigure}[b]{0.45\textwidth}
        \centering
        \includegraphics[width=\textwidth]{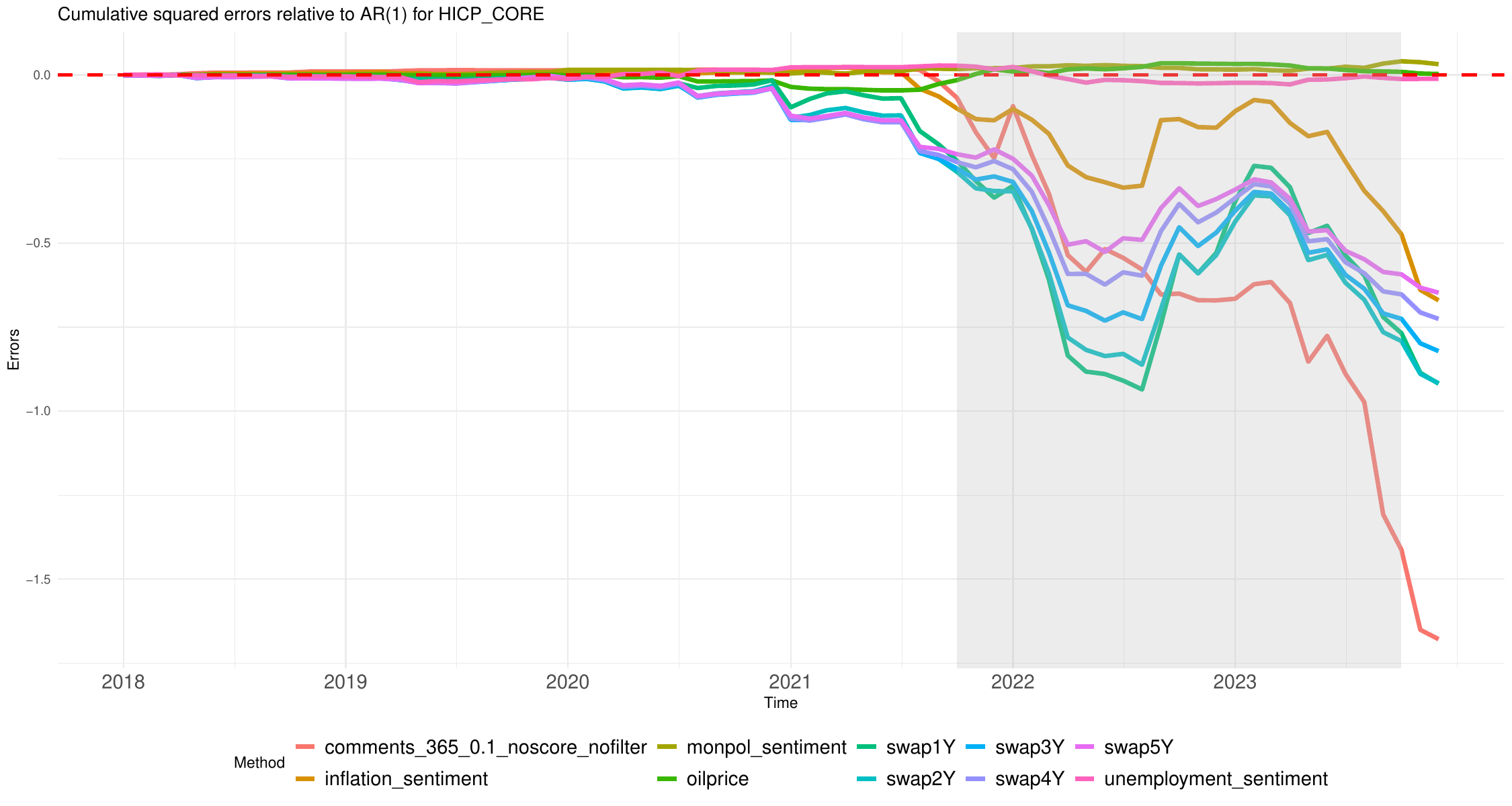}
        \caption{Core inflation}
        \label{fig:plot1}
    \end{subfigure}
    \hfill
    \begin{subfigure}[b]{0.45\textwidth}
        \centering
        \includegraphics[width=\textwidth]{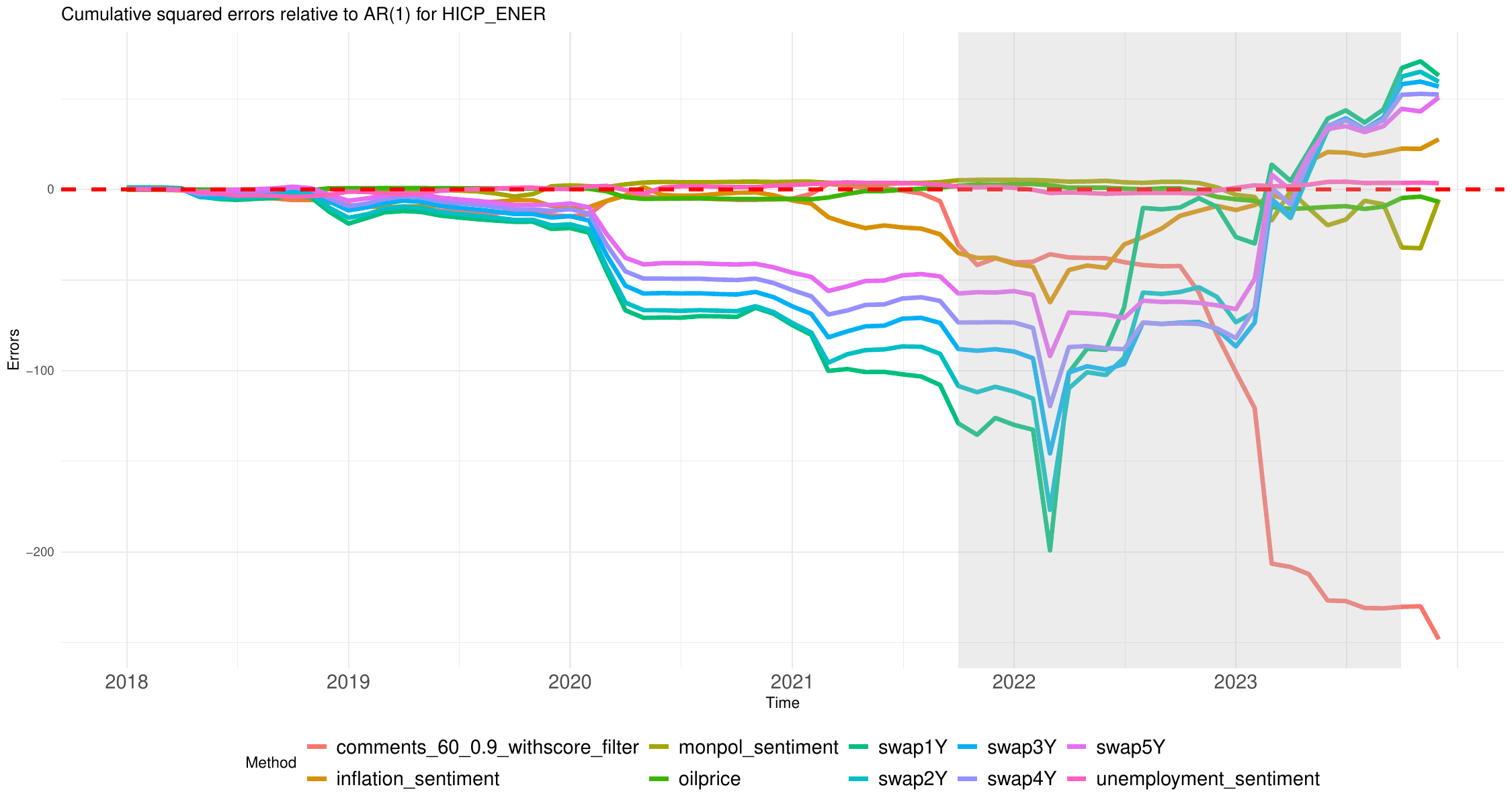}
        \caption{Energy price inflation}
        \label{fig:plot2}
    \end{subfigure}
    
    \vspace{0.5cm} 

    \begin{subfigure}[b]{0.45\textwidth}
        \centering
        \includegraphics[width=\textwidth]{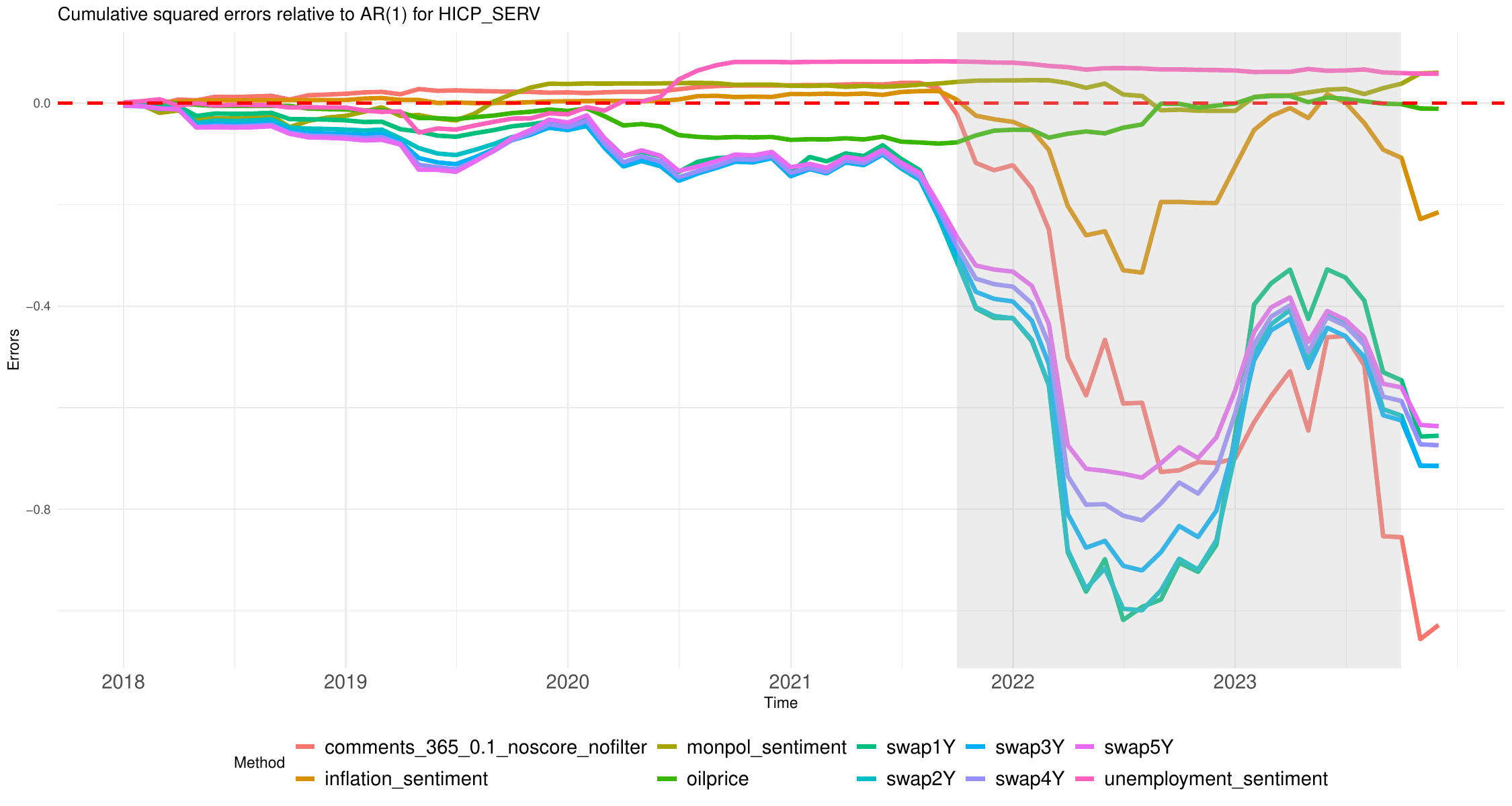}
        \caption{Service price inflation}
        \label{fig:plot3}
    \end{subfigure}
    \hfill
    \begin{subfigure}[b]{0.45\textwidth}
        \centering
        \includegraphics[width=\textwidth]{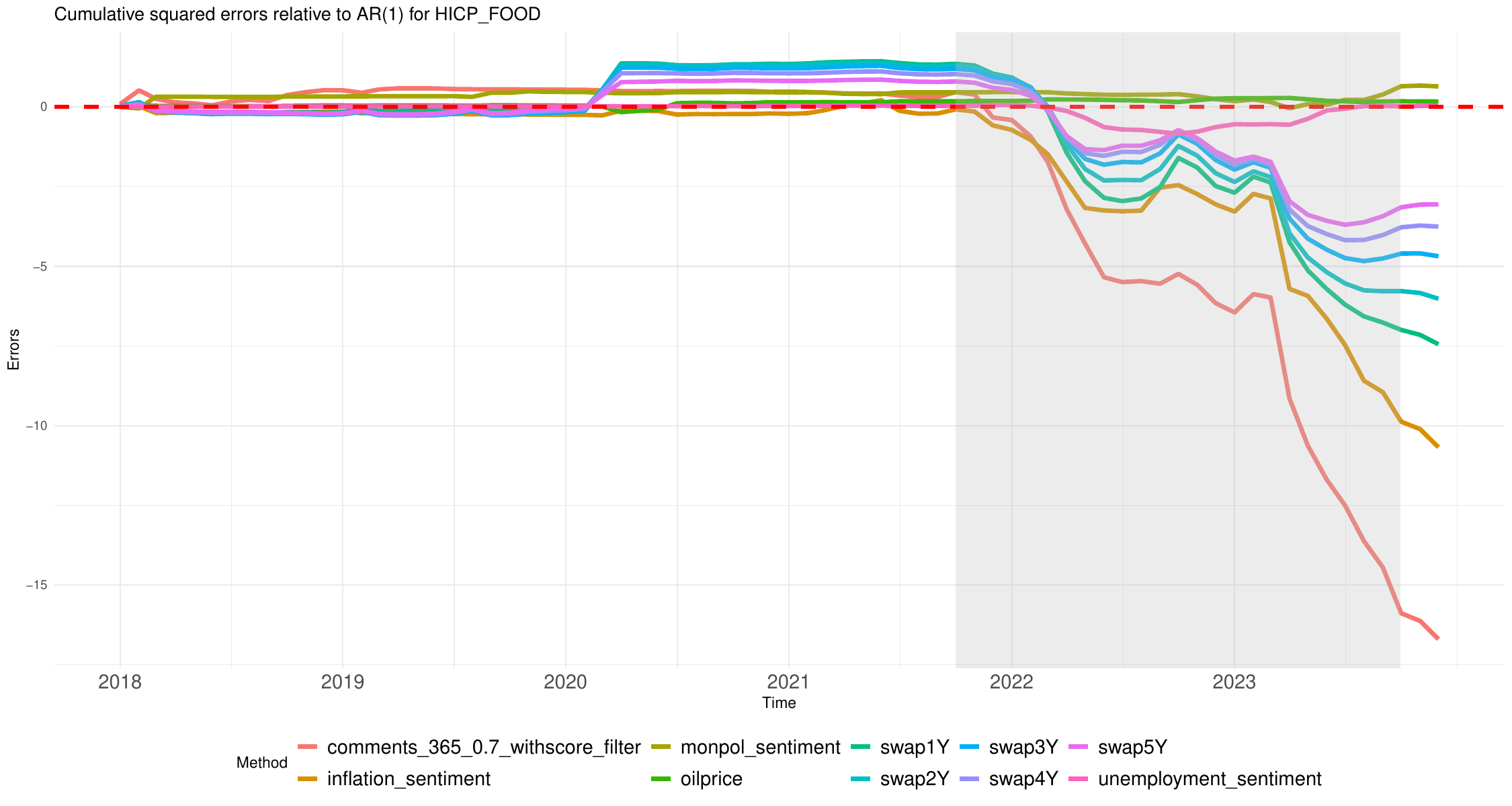}
        \caption{Food price inflation}
        \label{fig:plot4}
    \end{subfigure}
    
    \vspace{0.5cm} 

    \begin{subfigure}[b]{0.45\textwidth}
        \centering
        \includegraphics[width=\textwidth]{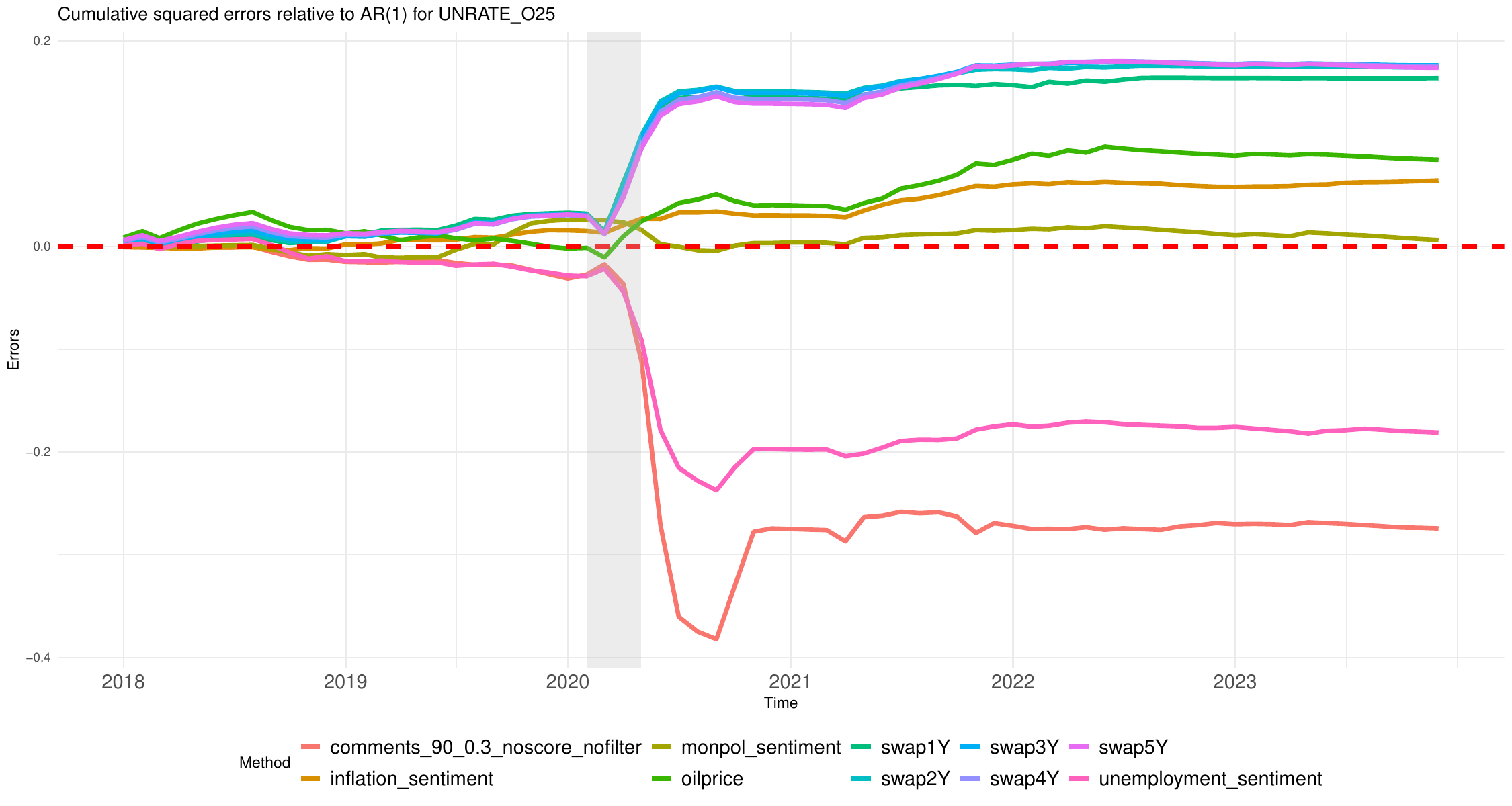}
        \caption{Unemployment $>$25}
        \label{fig:plot5}
    \end{subfigure}
    \hfill
    \begin{subfigure}[b]{0.45\textwidth}
        \centering
        \includegraphics[width=\textwidth]{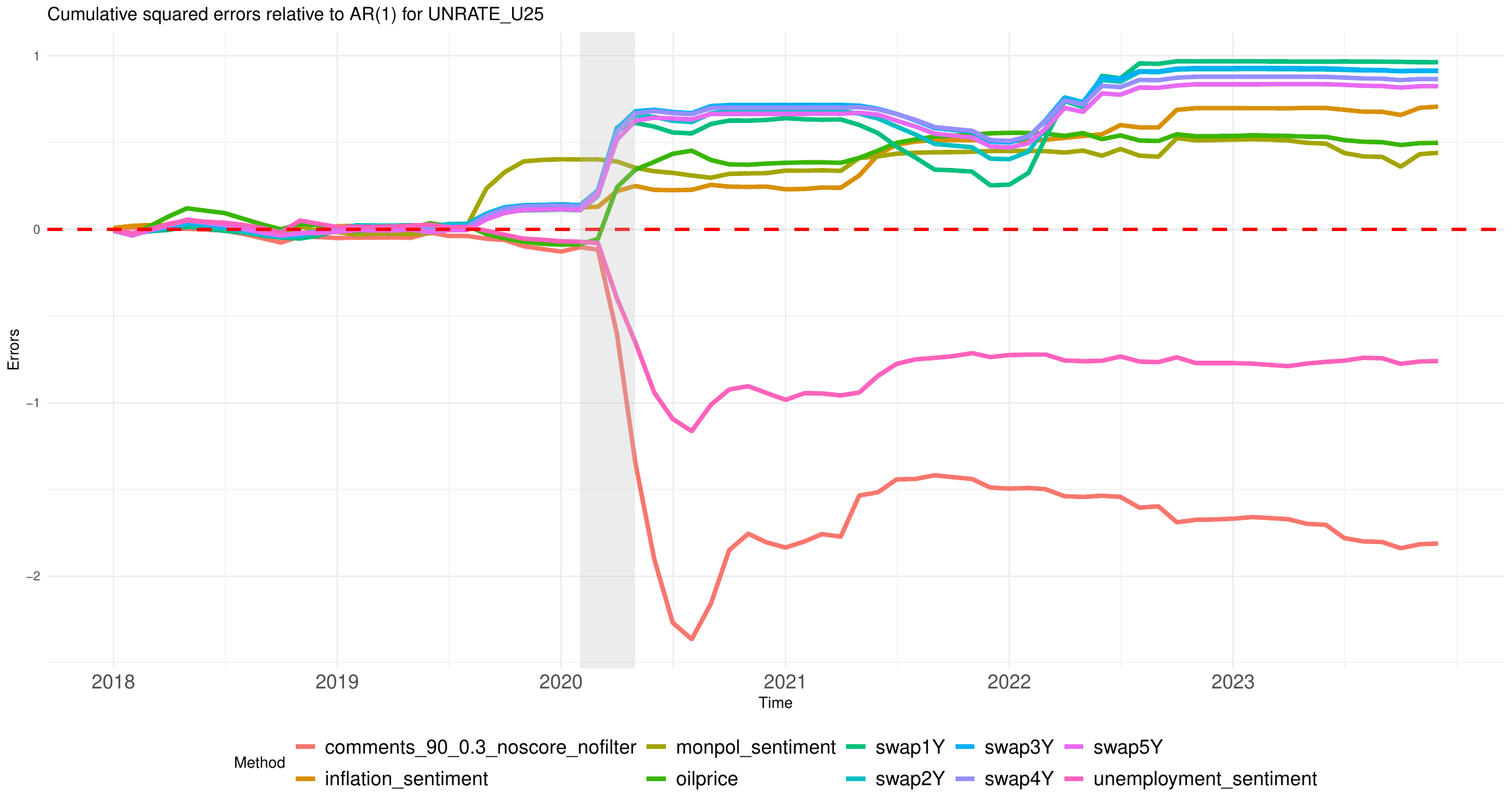}
        \caption{Unemployment $<$25}
        \label{fig:plot6}
    \end{subfigure}
    \caption{Cumulative squared errors of MIDAS models minus cumulative squared errors of AR(1) over time. Values below 0 mean better performance than the AR(1). Shaded areas are periods when HICP inflation was above 4\% and the COVID-19 recession months respectively. }
    \label{fig:more_horse_race}
\end{figure}

Figure \ref{fig:more_horse_race} presents the cumulated squared errors in differences to the AR(1) benchmark for the components of inflation and the unemployment rate. For all target variables, the Reddit signal involving social interaction produces the best result at the end of the sample. For inflation, the swap series are useful nowcasting predictors on the onset of the high inflation period, but less so at the end. For the unemployment rate components, the only real competitor to our Reddit series is the unemployment sentiment of \cite{barbaglia2024forecasting}.

\subsubsection{Nowcasting gains stability tests}\label{Appendix:GiacominiRossi}
\begin{figure}[H]
    \centering
    \begin{subfigure}[b]{0.75\textwidth}
        \centering
        \includegraphics[width=\textwidth]{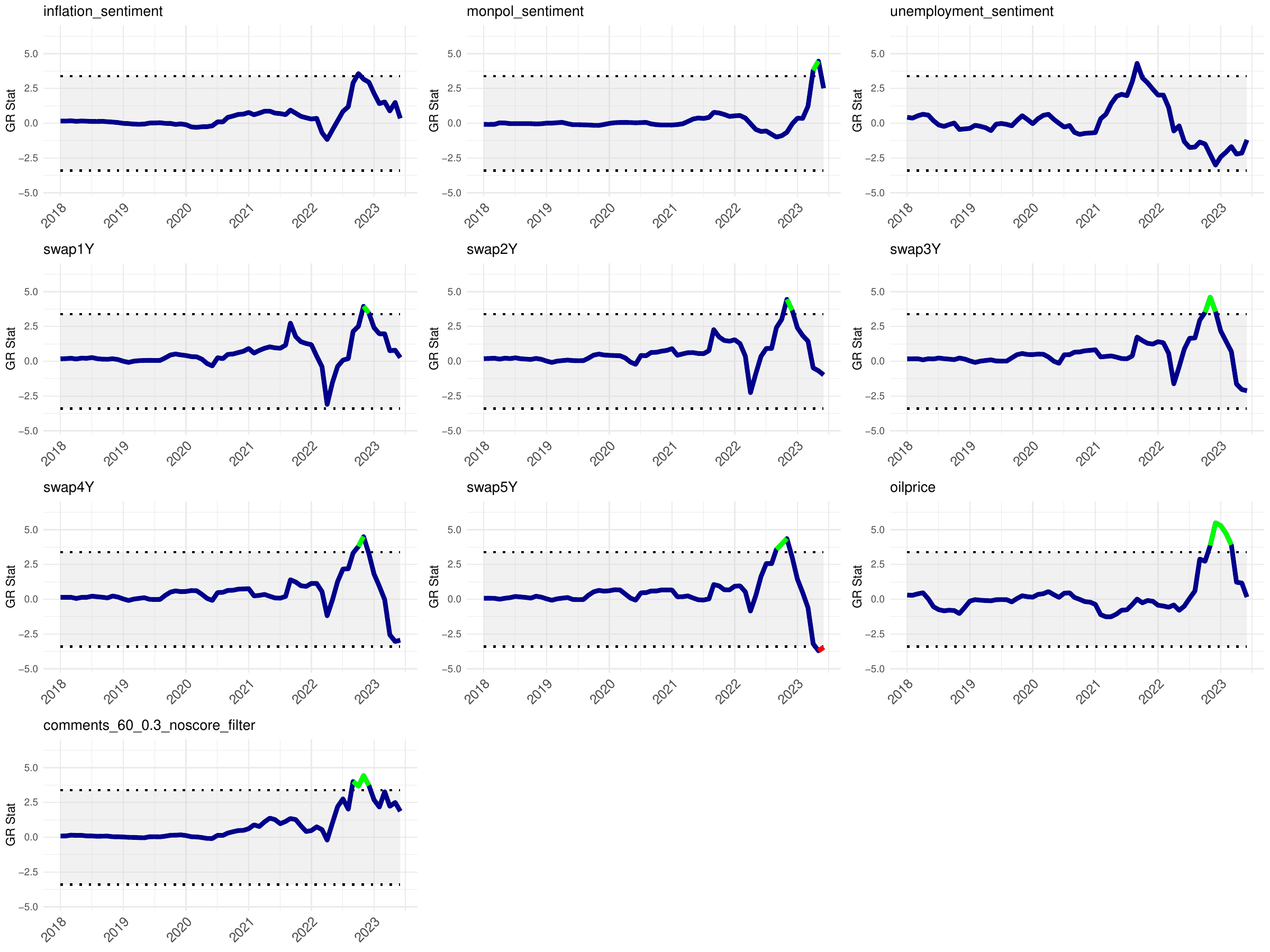}
        \caption{HICP inflation}
        \label{fig:plot1}
    \end{subfigure}
    \hfill    
    \begin{subfigure}[b]{0.75\textwidth}
        \centering
        \includegraphics[width=\textwidth]{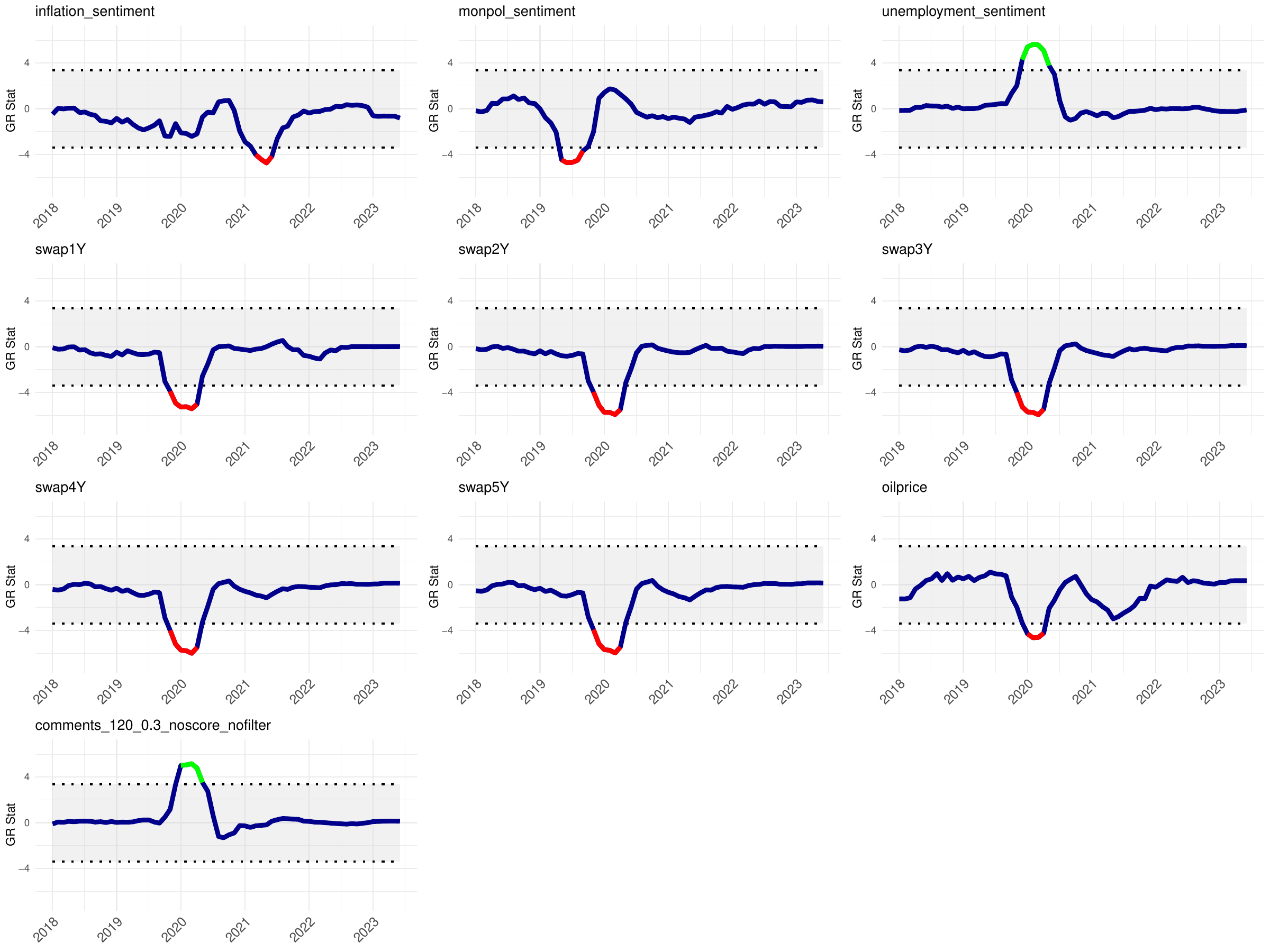}
        \caption{Unemployment rate}
        \label{fig:plot2}
    \end{subfigure}
    \hfill   
    \caption{Giacomini-Rossi test statistics and critical values over time. If the statistic is within the bounds there is no evidence of significantly different performance. If the statistic exceeds the upper bound (green chunks) the model outperforms the AR(1), otherwise (red chunks) it underperforms. }\label{fig:giacomini-rossi}
\end{figure}

Given that the performance gains appear to be driven by unusual times, we check the models' performance using the instability test of \cite{giacomini2010forecast}, where we choose the moving-window in the evaluation sample to be of size 10\% ($\mu = 0.1$). The test provides evidence on the time periods in which the the MIDAS specification achieves significant gains (or sustains losses) compared to the AR(1) benchmark. Figure \ref{fig:giacomini-rossi} shows that, indeed, the Reddit signal for inflation is significantly better than the benchmark during the hyperinflation period around the end of 2022. In the case of the unemployment rate, the indicator provides improvements over the benchmark in the COVID-19 period. \\

\subsubsection{The role of daily information flow}\label{Appendix:DailyInforamtion}
To better judge the real-time value of the daily Reddit series, we conduct an additional check. We relax the assumption that the daily information is used once all of it has been revealed at the end of the target period. Instead, we assume that the information cutoff occurs 7, 14, 21, and 28 days before the end of the period that is to be nowcasted. This resembles the nowcasting setup of \cite{banbura2013now}, who emulate the flow of information through the months of the quarter to be nowcasted. In principle, more information should improve the nowcasts, but a practitioner may not have the full month available at the point when the nowcast is to be released.

\begin{figure}[H]
    \centering
    \begin{subfigure}[b]{0.45\textwidth}
        \centering
        \includegraphics[width=\textwidth]{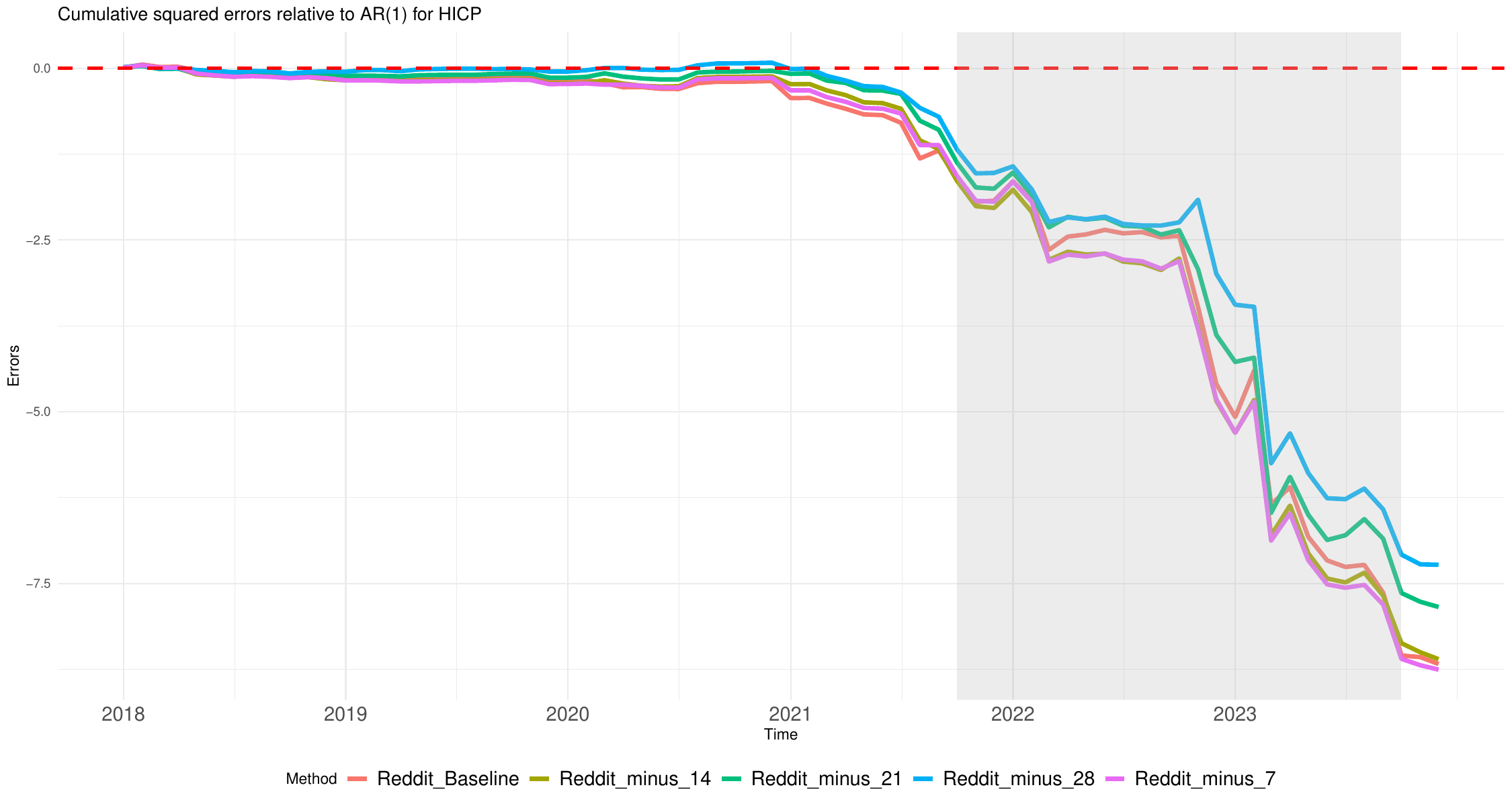}
        \caption{Cumulative Errors HICP inflation}
        \label{fig:plot1}
    \end{subfigure}
    \hfill
    \begin{subfigure}[b]{0.45\textwidth}
        \centering
        \includegraphics[width=\textwidth]{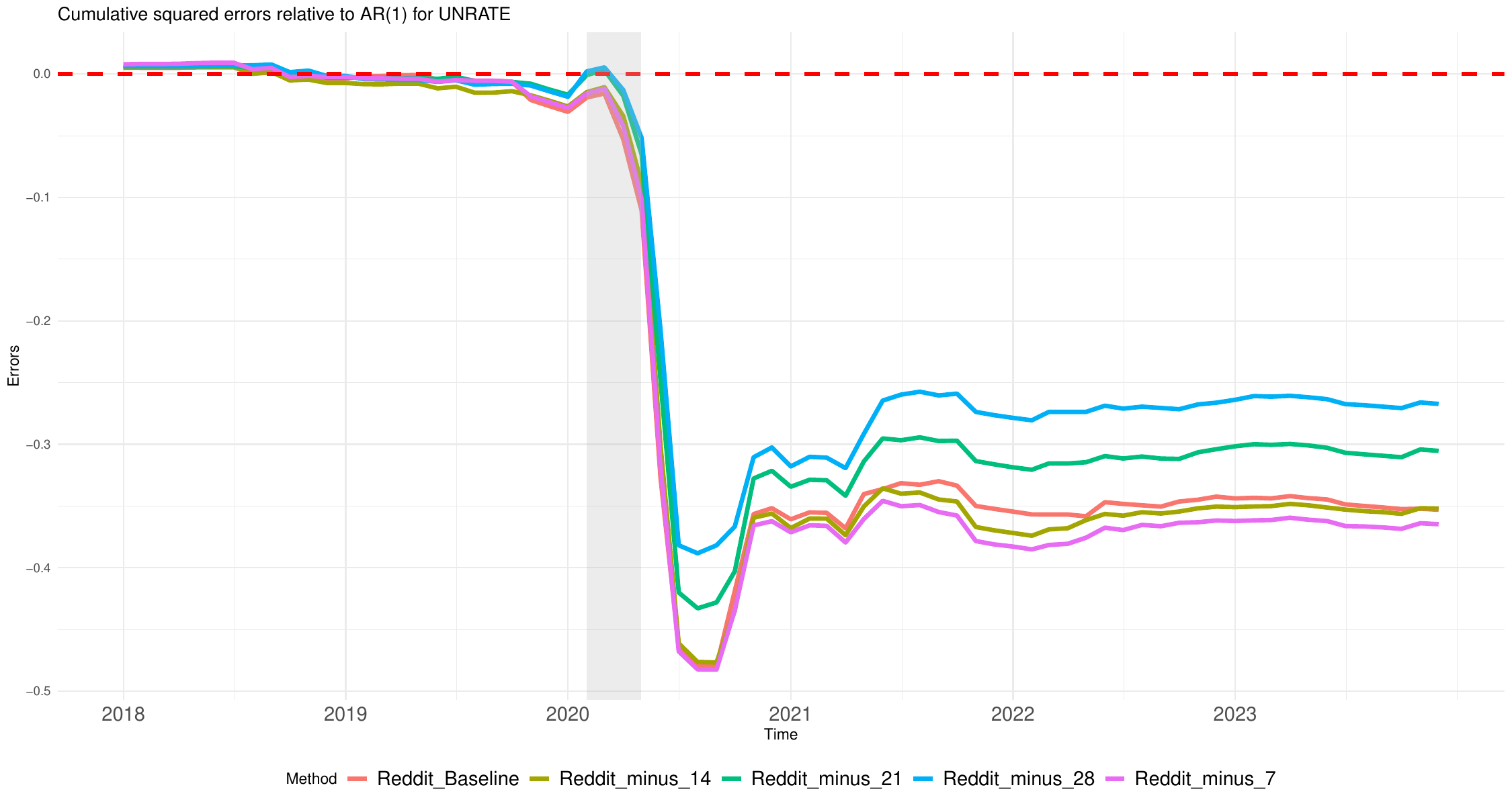}
        \caption{Cumulative Errors Unemployment rate}
        \label{fig:plot2}
    \end{subfigure}
    
    \vspace{0.5cm} 

    \begin{subfigure}[b]{0.45\textwidth}
        \centering
        \includegraphics[width=\textwidth]{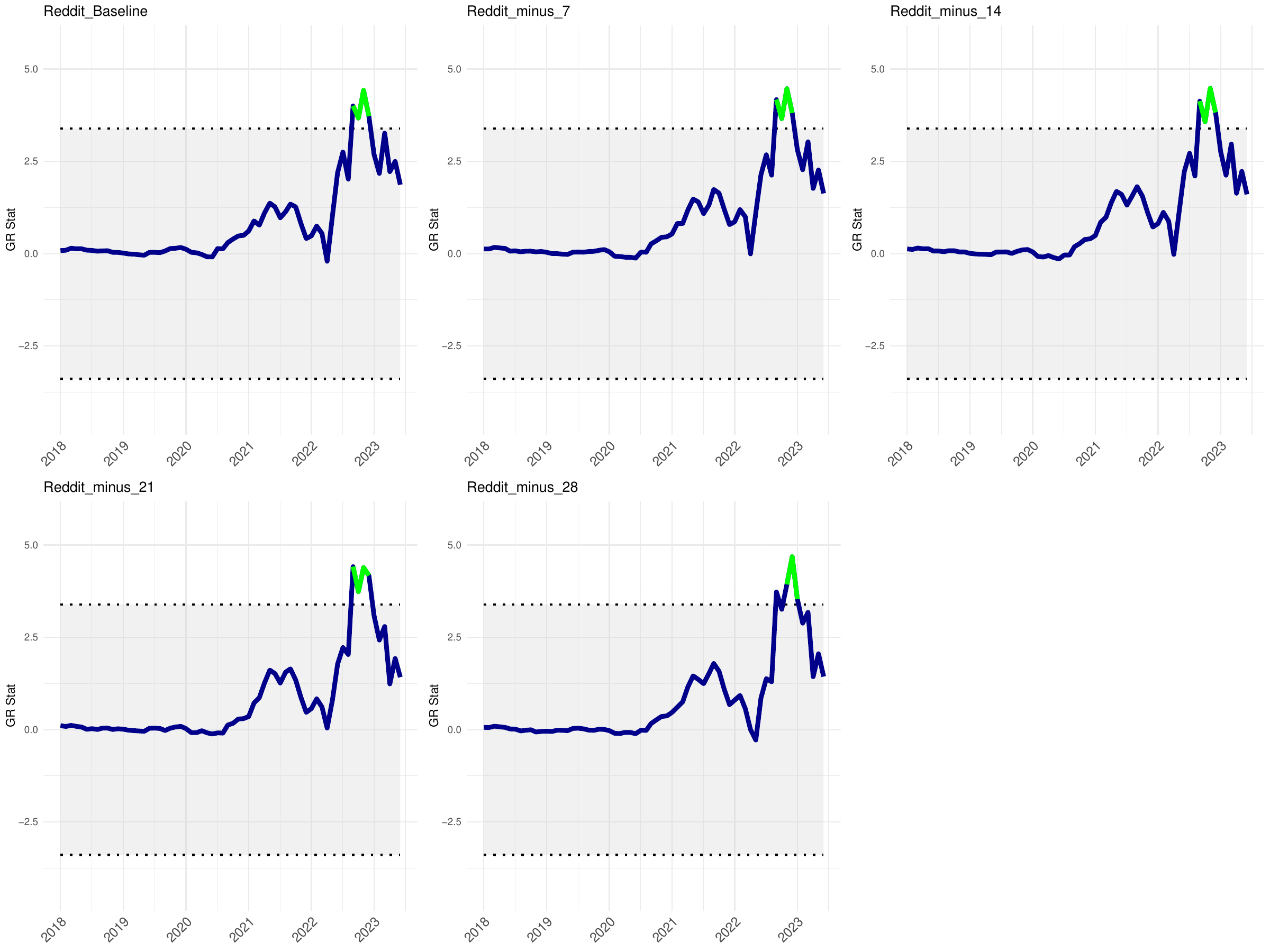}
        \caption{Giacomini-Rossi HICP inflation}
        \label{fig:plot3}
    \end{subfigure}
    \hfill
    \begin{subfigure}[b]{0.45\textwidth}
        \centering
        \includegraphics[width=\textwidth]{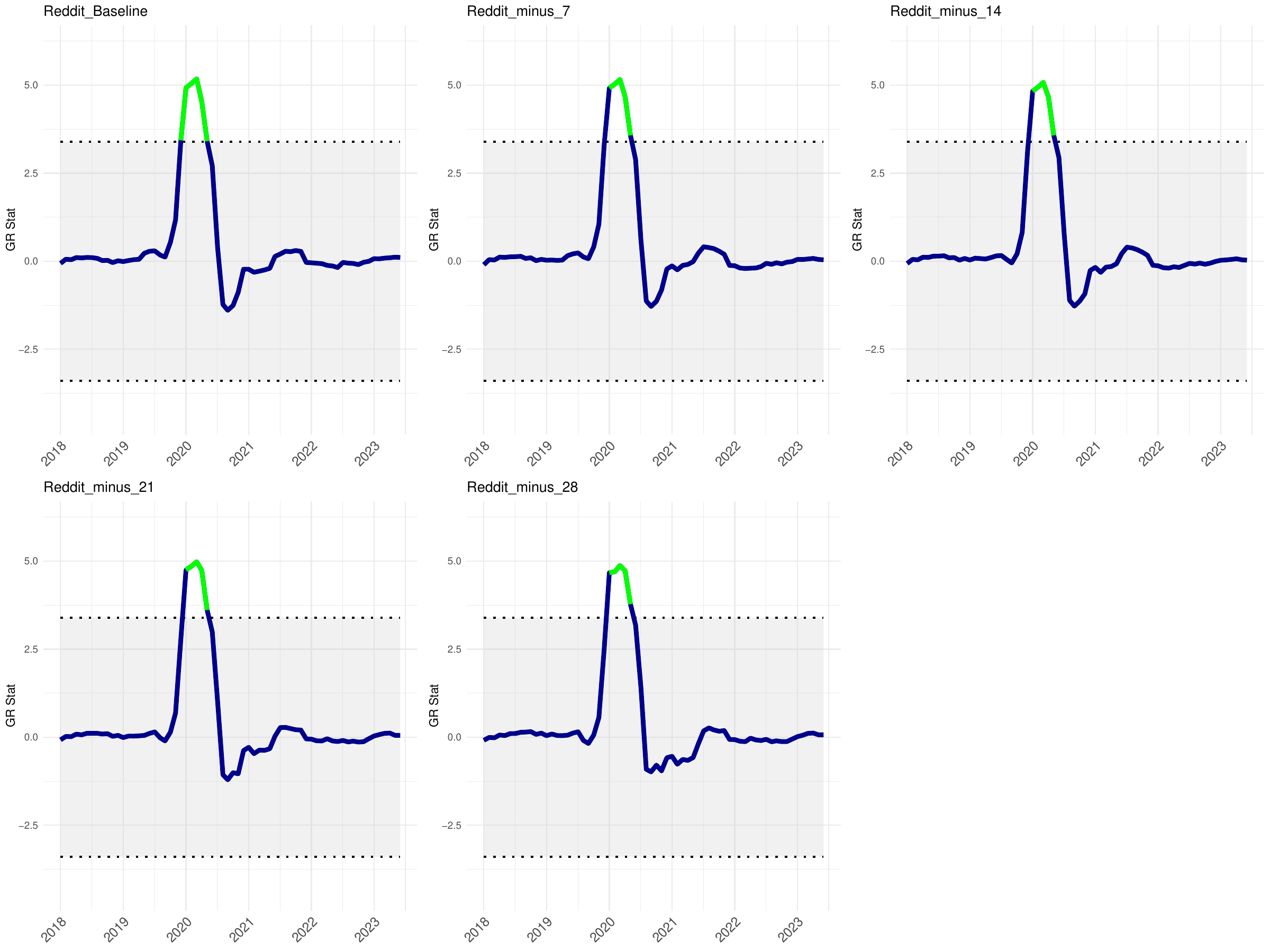}
        \caption{Giacomini-Rossi Unemployment rate}
        \label{fig:plot4}
    \end{subfigure}
    \caption{Cumulative squared errors of MIDAS with information cutoffs 0,7,14,21, and 28 days before the end of the nowcast period. Values are calculated subtracting cumulative squared errors of AR(1) over time. Values below 0 mean better performance than the AR(1). Shaded areas are periods when HICP inflation was above 4\% and the COVID-19 recession months respectively. Giacomini-Rossi test statistics and critical values over time. If the statistic is within the bounds there is no evidence of significantly different performance. If the statistic exceeds the upper bound (green chunks) the model outperforms the AR(1), otherwise (red chunks) it underperforms. }
    \label{fig:infoflow}
\end{figure}

Figure \ref{fig:infoflow} reports the result of the MIDAS models for the best performing Reddit specifications (in terms of RMSFE) for HICP inflation and the unemployment rate with varying information cutoffs. For both variables, there is some benefit to considering additional information, but this is clear up to around 14 days before the cutoff only. Thereafter performances are essentially equal. The stability tests emphasize that the information contained in the signal is significantly useful in all cases, highlighting the real time value of the Reddit data.

\subsubsection{The role of smoothing}\label{Appendix:Smoothing}
The overall horse race results reported above suggest that unemployment nowcasts always favor a smoothing window of 90 days. For the inflation series the preferred window sizes are 60 and 365 days for the five targets. To better understand the influence of the MA smoothing window, we plot the relative nowcasting performance for MIDAS models which only use submissions and MIDAS models which use submissions and comments over different window sizes, always scored against the AR(1) benchmark for ease of comparison. 

\begin{figure}[H]
    \centering
    \begin{subfigure}[b]{0.45\textwidth}
        \centering
        \includegraphics[width=\textwidth]{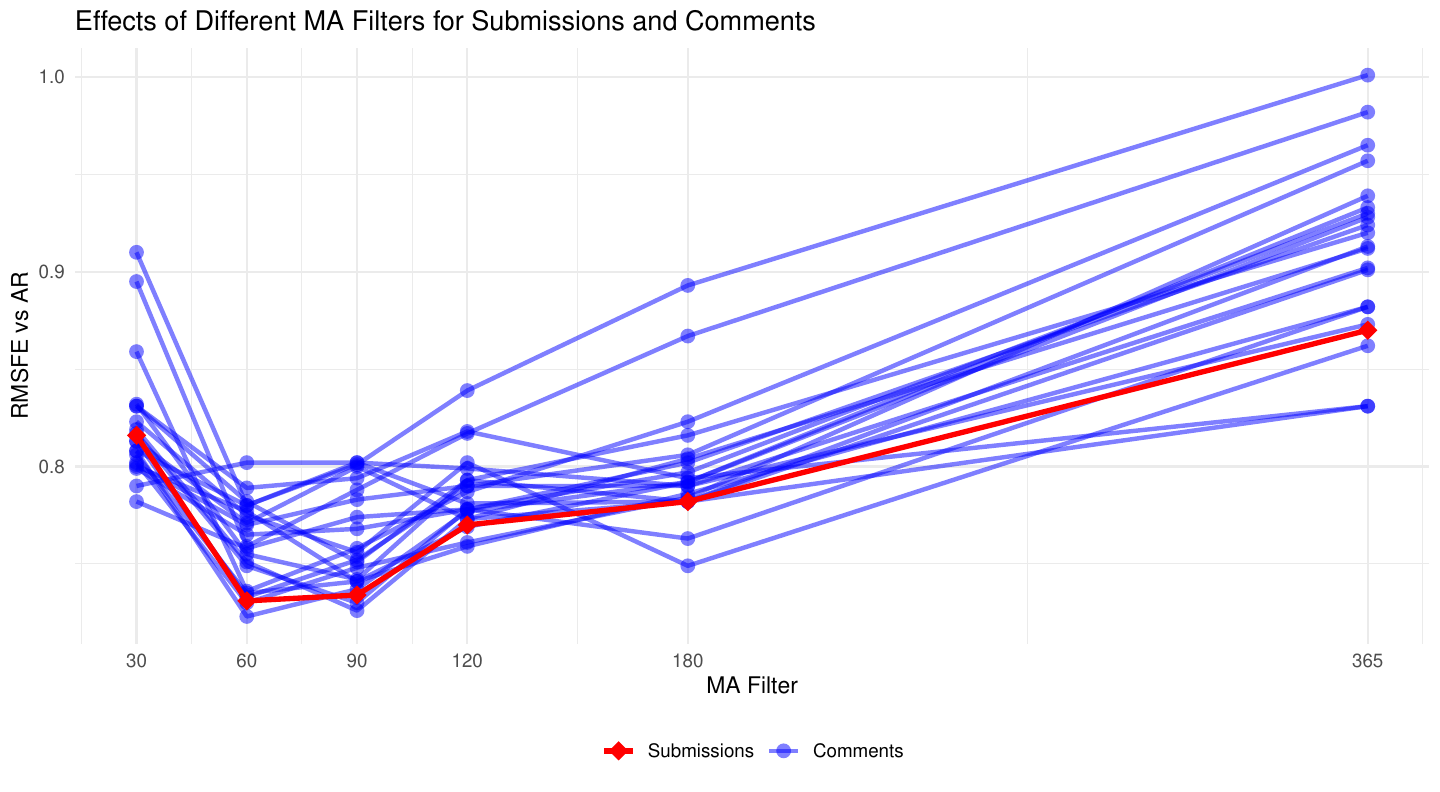}
        \caption{HICP inflation}
        \label{fig:plot1}
    \end{subfigure}
    \hfill
    \begin{subfigure}[b]{0.45\textwidth}
        \centering
        \includegraphics[width=\textwidth]{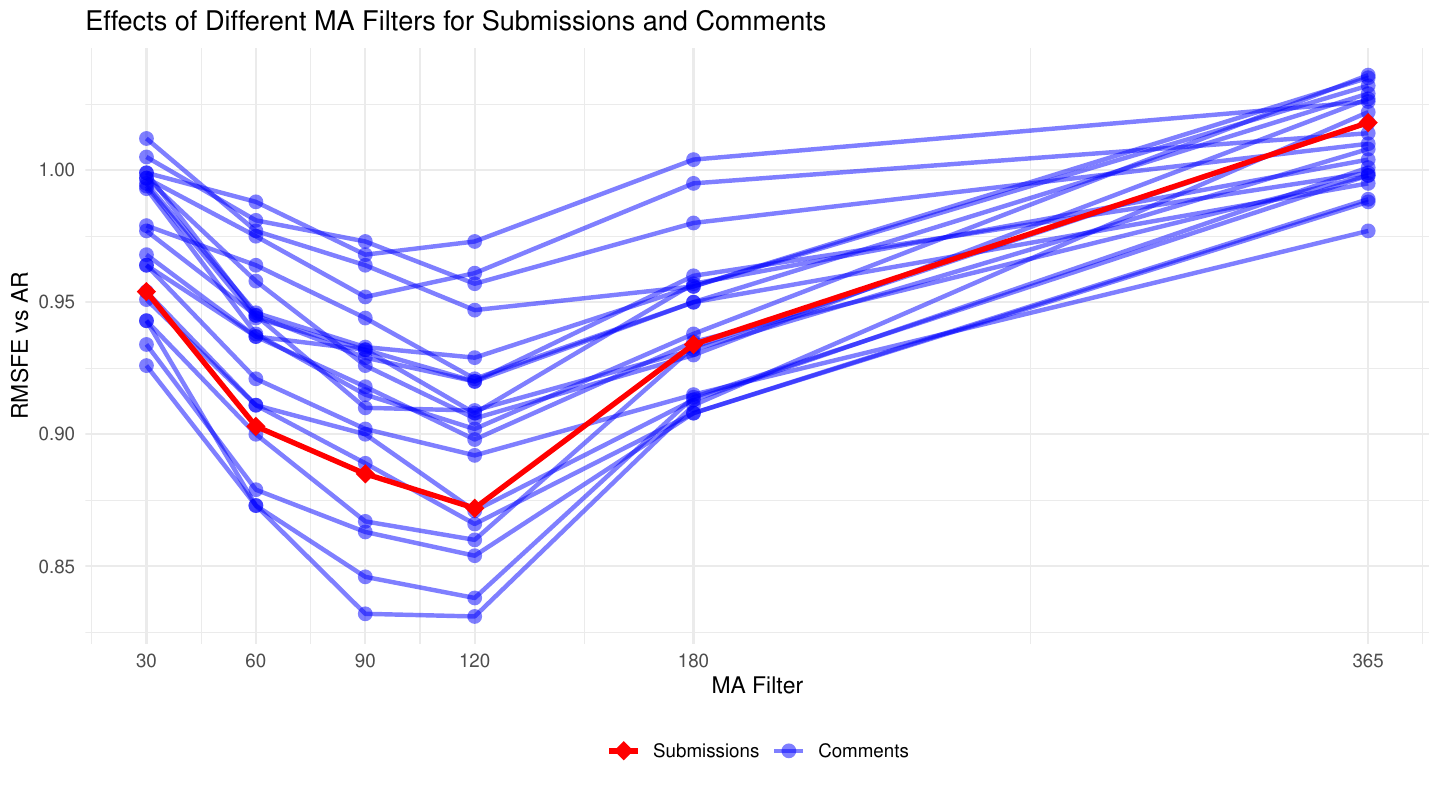}
        \caption{Unemployment rate}
        \label{fig:plot2}
    \end{subfigure}
    
    \vspace{0.5cm} 

    \begin{subfigure}[b]{0.45\textwidth}
        \centering
        \includegraphics[width=\textwidth]{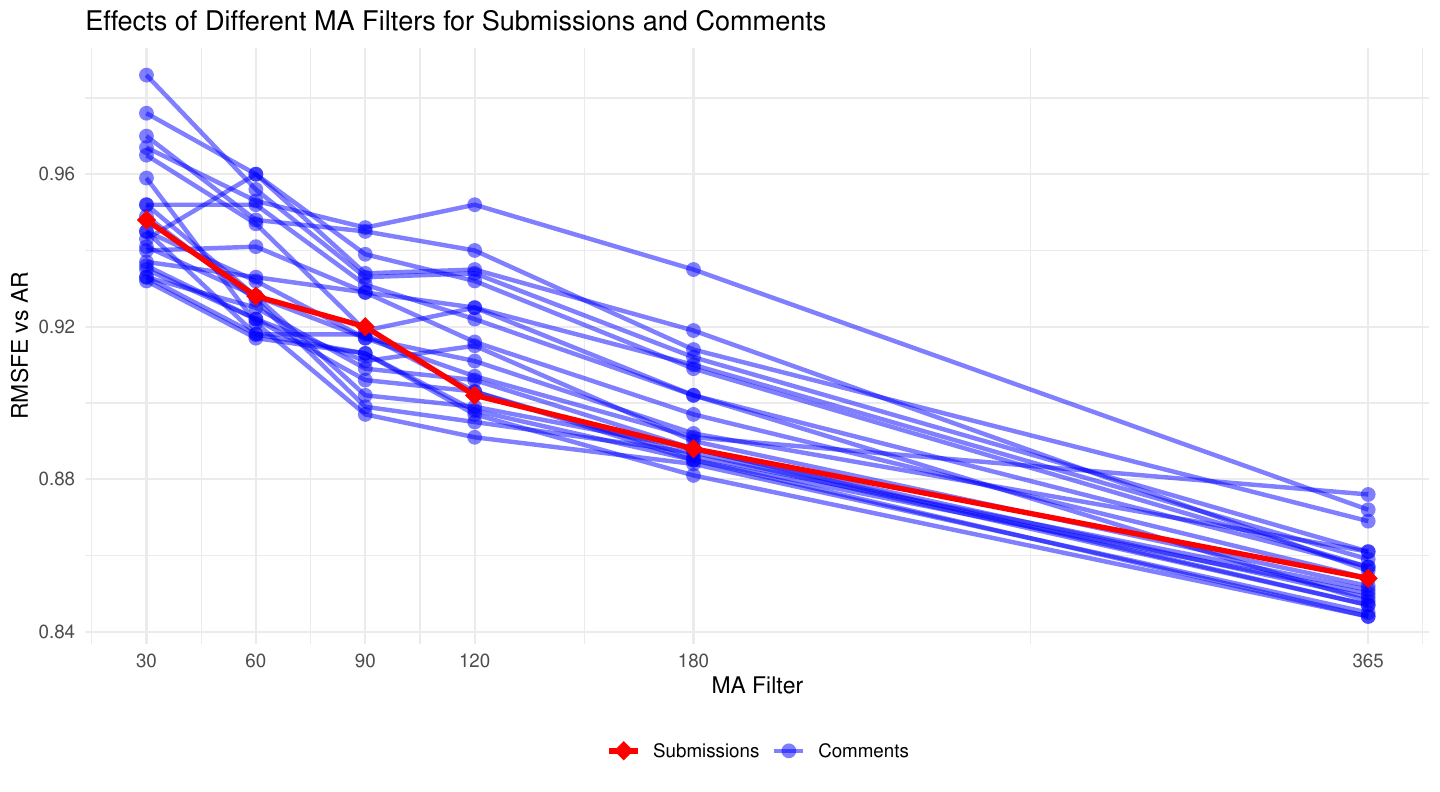}
        \caption{HICP Core}
        \label{fig:plot3}
    \end{subfigure}
    \hfill
    \begin{subfigure}[b]{0.45\textwidth}
        \centering
        \includegraphics[width=\textwidth]{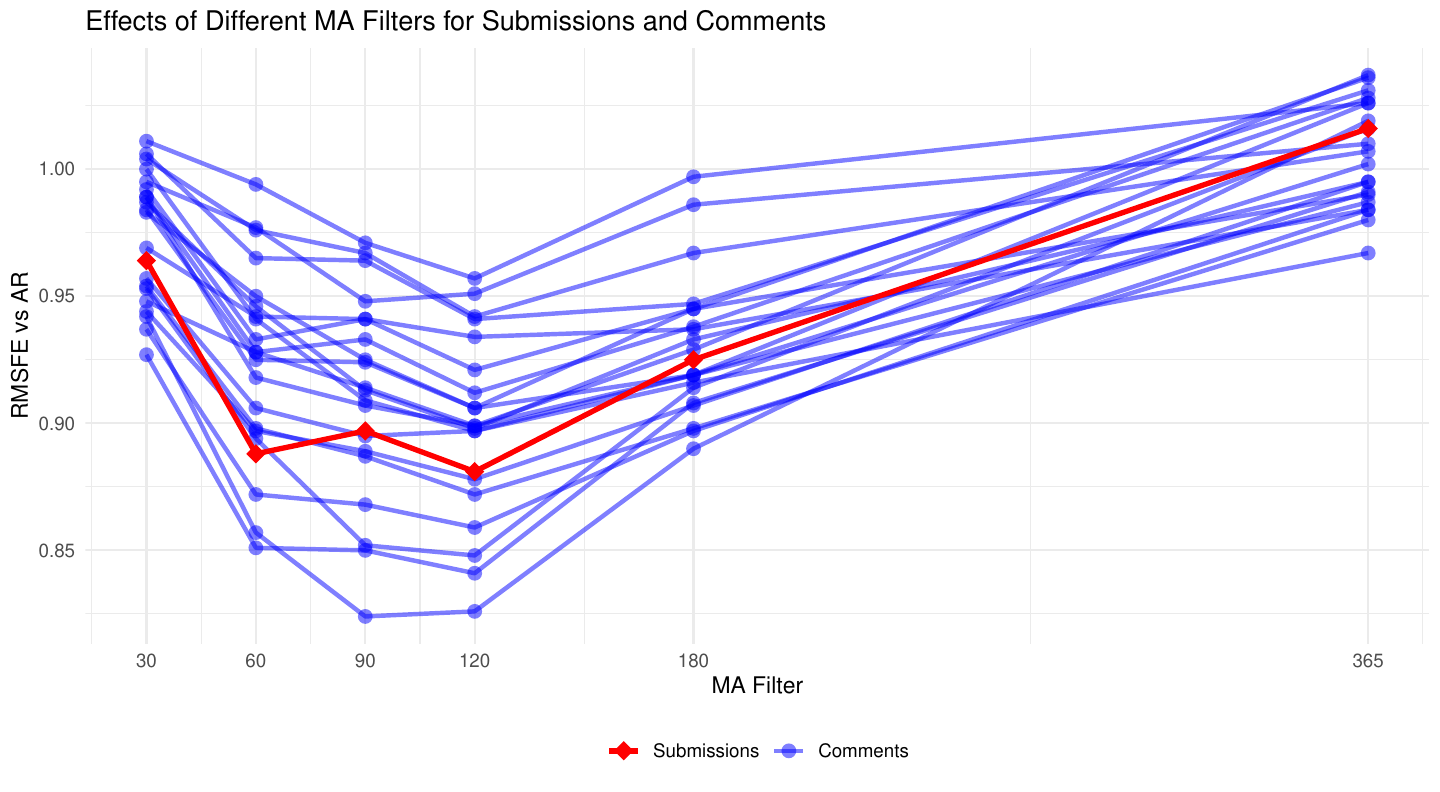}
        \caption{Unemployment $>$25}
        \label{fig:plot4}
    \end{subfigure}
    
    \vspace{0.5cm} 

    \begin{subfigure}[b]{0.45\textwidth}
        \centering
        \includegraphics[width=\textwidth]{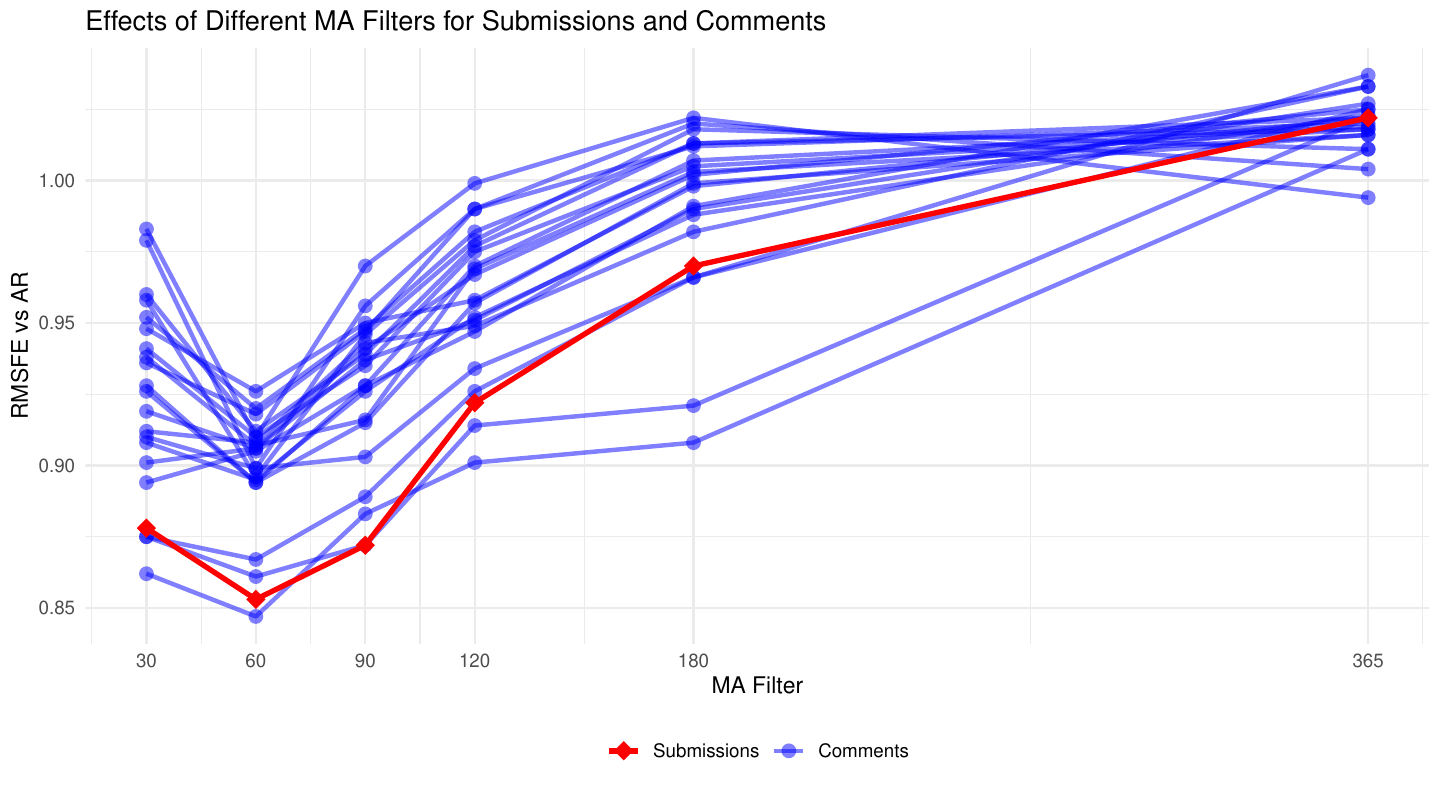}
        \caption{HICP Energy}
        \label{fig:plot5}
    \end{subfigure}
    \hfill
    \begin{subfigure}[b]{0.45\textwidth}
        \centering
        \includegraphics[width=\textwidth]{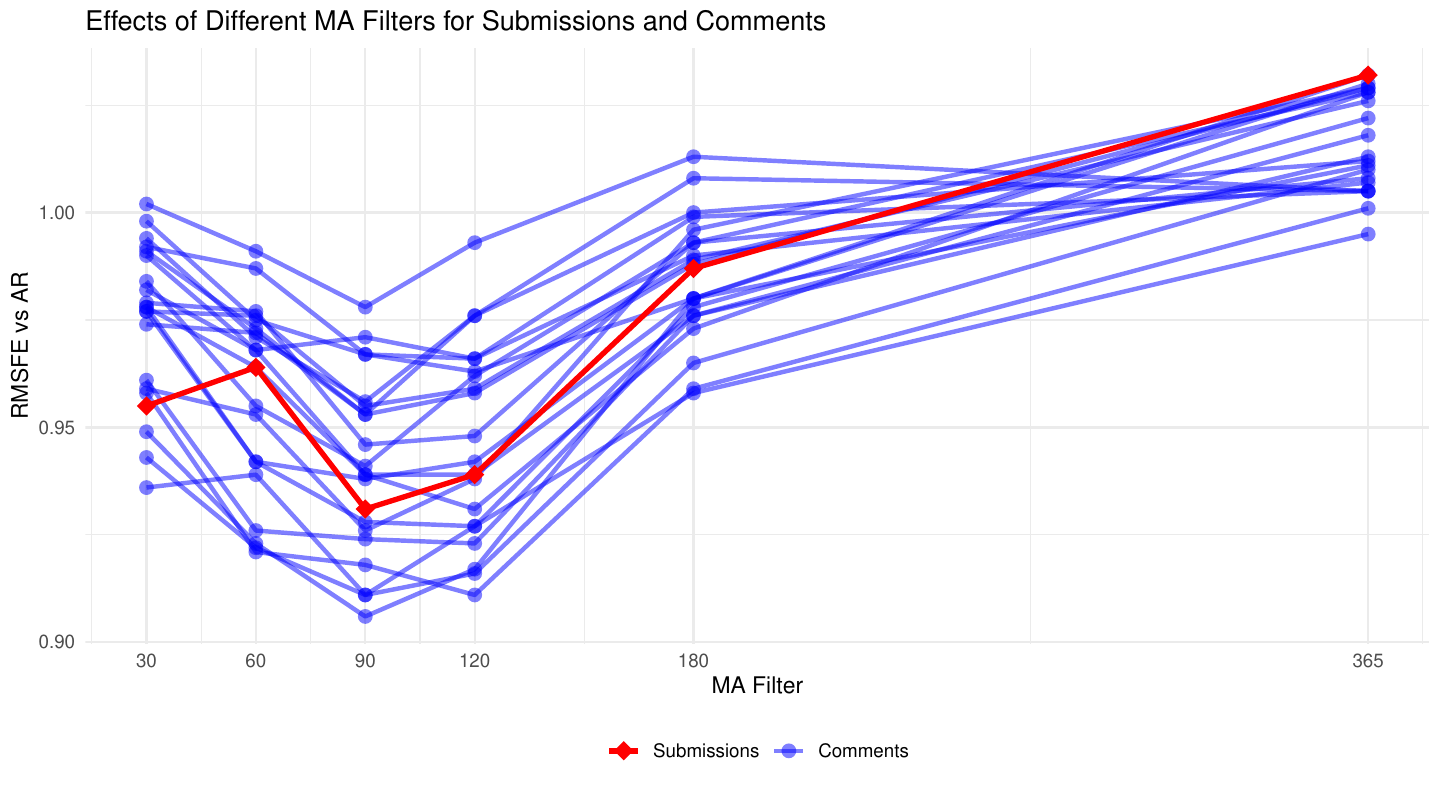}
        \caption{Unemployment $<$25}
        \label{fig:plot6}
    \end{subfigure}
    
    \vspace{0.5cm} 

    \begin{subfigure}[b]{0.45\textwidth}
        \centering
        \includegraphics[width=\textwidth]{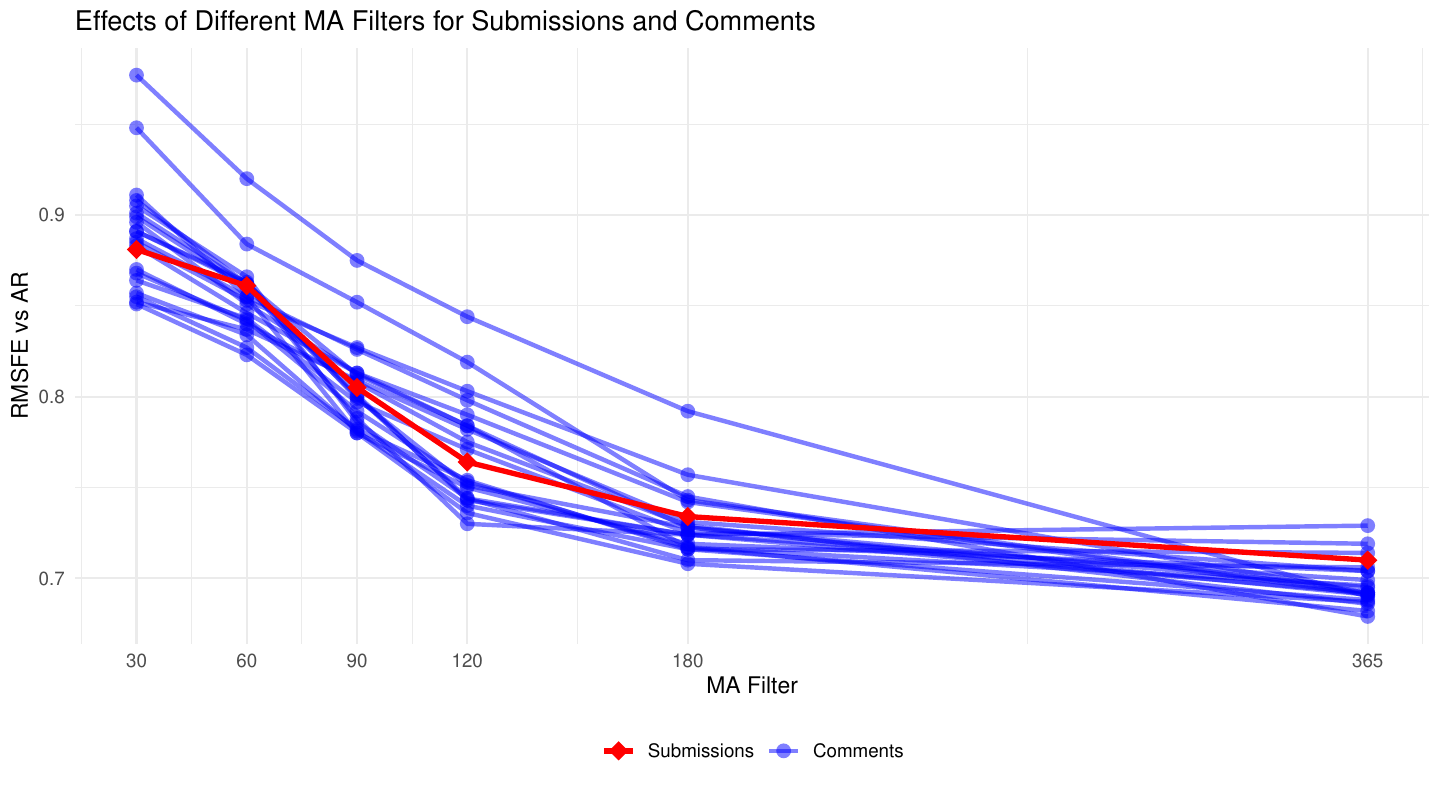}
        \caption{HICP Food}
        \label{fig:plot7}
    \end{subfigure}
    \hfill
    \begin{subfigure}[b]{0.45\textwidth}
        \centering
        \includegraphics[width=\textwidth]{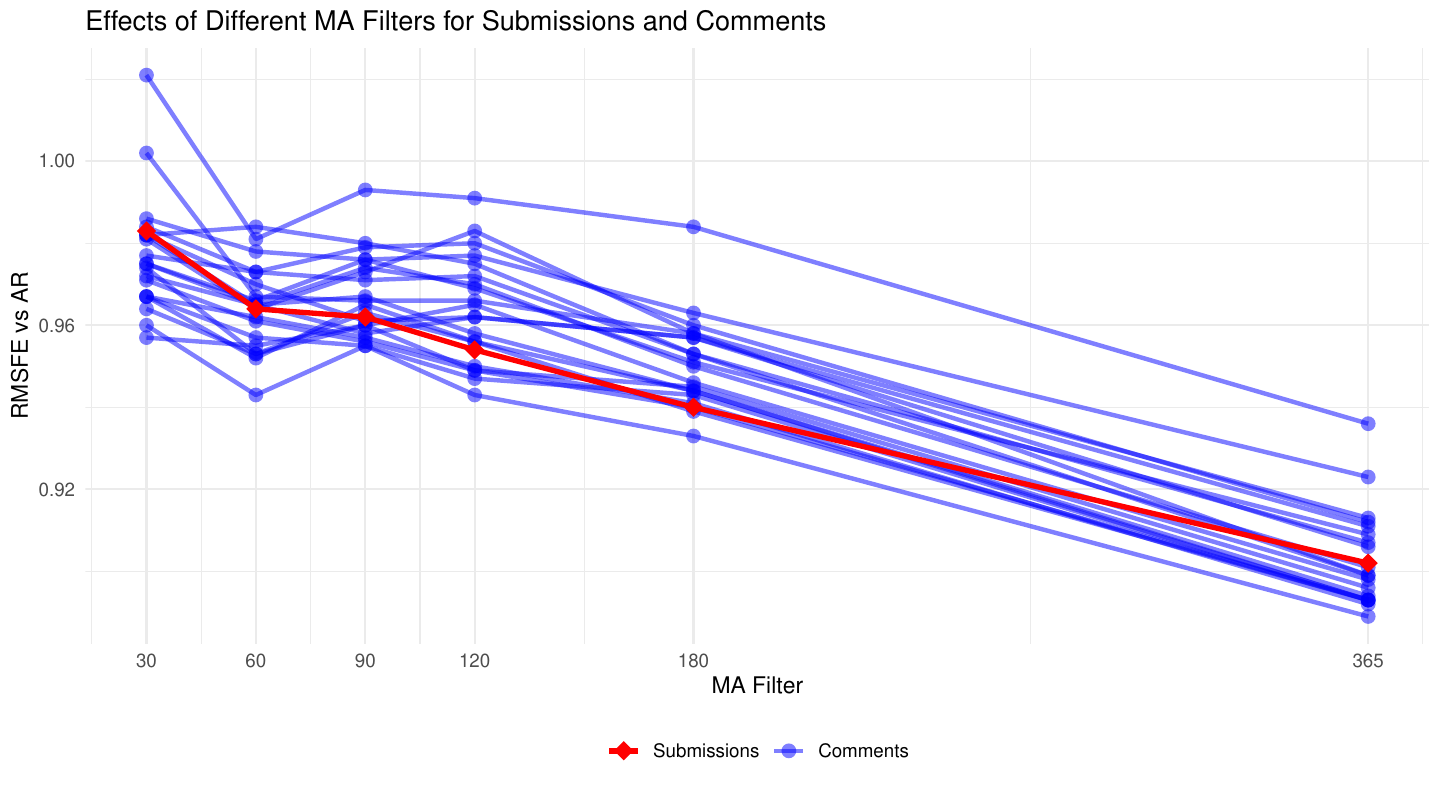}
        \caption{HICP Services}
        \label{fig:plot8}
    \end{subfigure}

    \caption{RMSFE scores for Reddit signals against AR(1) benchmark compared for different sizes of the MA filter window in days.}
    \label{fig:mafilters_lines}
\end{figure}

Figure \ref{fig:mafilters_lines} clearly shows that for HICP inflation Reddit signals, the optimal smoothing window width is around 60 to 90 days, whereas for the unemployment rate the optimum is between 90 and 120 days. The picture is very similar for energy price inflation which has its optimum at short horizons, as it is more volatile. In the case of the other HICP components, there is no clear kink and the optimal smoothing window size is the maximum we have tested (365 days). In some cases (core, services) it appears as if even wider windows could be beneficial, in others the plot flattens between 180 and 365 days (food). 

\subsubsection{The role of thresholds and pre-filtering for comments}\label{Appendix:SocialInteraction}
While the choice of smoothing window is clearly important in all plots there is at least one specification involving comments that beats the optimal submissions-only model, for the unemployment rate this is more obvious than for the inflation data. To better understand what components -- if any -- of the comments signals (smoothing, thresholds, scoring, or pre-filtering) contribute systematically to these performance gains, we break the RMSFE scores against the AR(1) down by these four dimensions. 

\begin{figure}[H]
    \centering
    \begin{subfigure}[b]{0.9\textwidth}
        \centering
        \includegraphics[width=\textwidth]{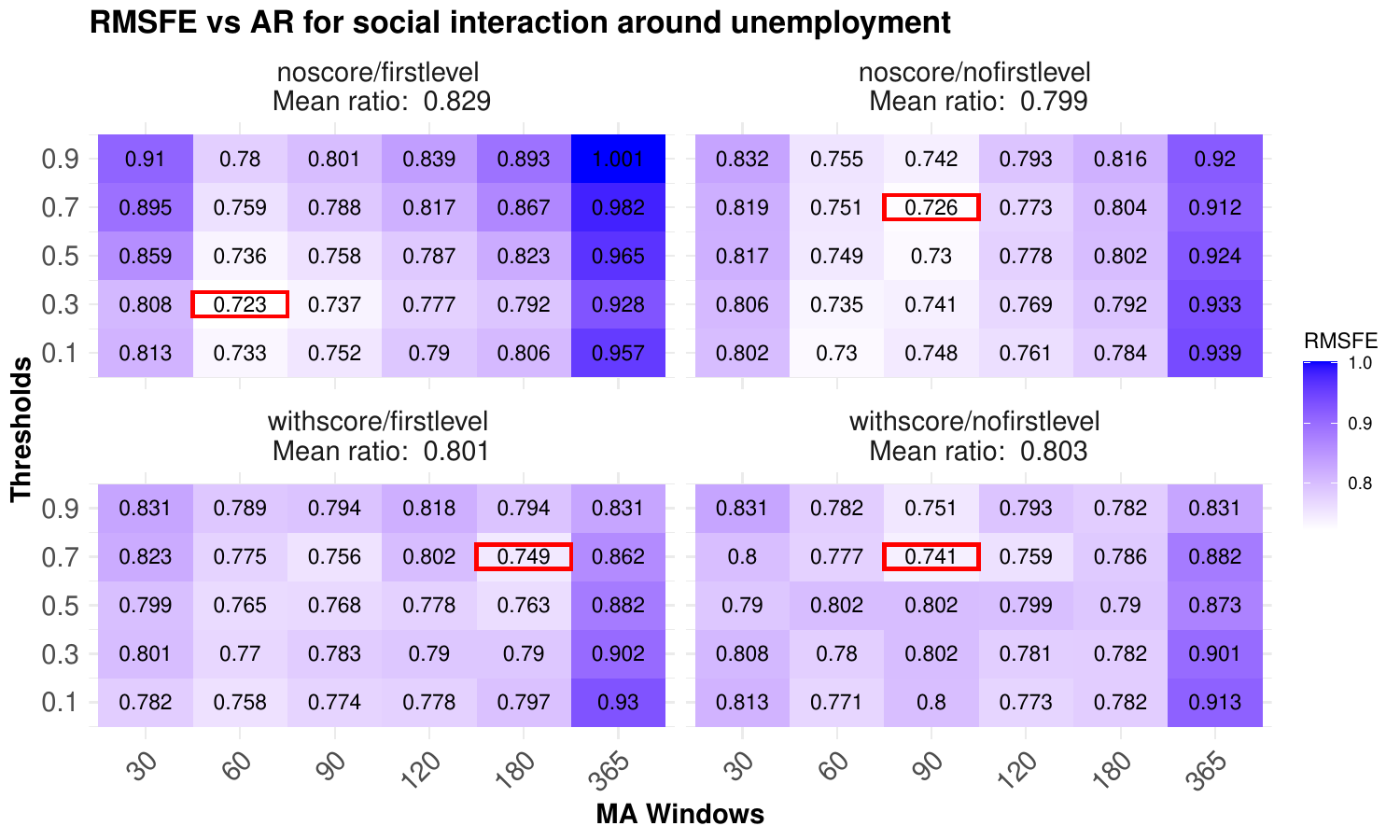}
        \caption{HICP inflation}
        \label{fig:plot1}
    \end{subfigure}
    \hfill    
    \begin{subfigure}[b]{0.9\textwidth}
        \centering
        \includegraphics[width=\textwidth]{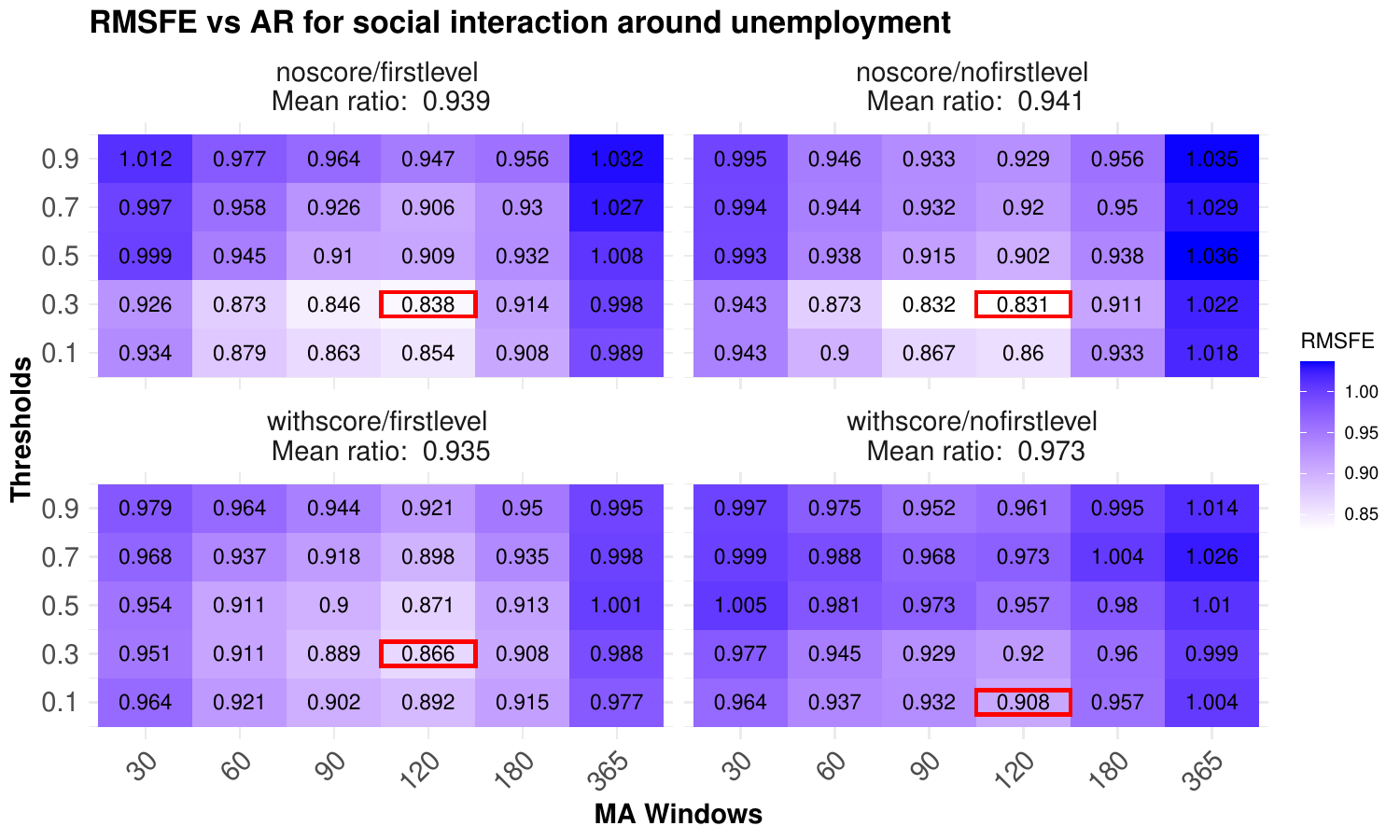}
        \caption{Unemployment rate}
        \label{fig:plot2}
    \end{subfigure}
    \hfill   
    \caption{Relative RMSFE scores against an AR(1) for MIDAS specifications estimated with different variations of Reddit comments signals. Brighter colors imply better scores, values below 1 imply that the Reddit signal improves upon the AR(1). }\label{fig:heatmaps}
\end{figure}

The results of this are displayed in Figure \ref{fig:heatmaps}. For HICP inflation we see that there is no clear pattern that would indicate significant gains stemming from one of the four dimensions that influence the construction of the Reddit signals. Using the scoring rule based on upvotes and downvotes seems to produce slightly better results than the specification which does not incorporate this extra layer of information even though the overall best performing model does not apply it. The pre-filtering using the list of keywords does not appear to be relevant and there is no clear guidance on the value of the threshold which should optimally be employed. The results are clearer in the case of the unemployment rate. The pre-filtering improves performance, especially in combination with the upvote/downvote scoring, at least on average. Again, the standout performance is achieved by a specification which uses neither of these. A low threshold together with a smoothing window of 120 is optimal.

\end{document}